\documentclass[preprint]{emulateapj}

\newcommand{\etal}{et al.}
\newcommand{\jybeam}{Jy beam$^{-1}$}
\newcommand{\mjybeam}{mJy beam$^{-1}$}
\newcommand{\ntwohp}{N$_2$H$^+$ $J=1-0$}
\newcommand{\NTWOHP}{N$_2$H$^+$}
\newcommand{\co}{$^{12}$CO $J=1-0$}
\newcommand{\CO}{$^{12}$CO}
\newcommand{\ceighteeno}{C$^{18}$O $J=1-0$}
\newcommand{\CEIGHTEENO}{C$^{18}$O}
\newcommand{\thirteenco}{$^{13}$CO $J=1-0$}
\newcommand{\THIRTEENCO}{$^{13}$CO}
\newcommand{\hcop}{HCO$^+$ $J=1-0$}
\newcommand{\HCOP}{HCO$^+$}
\newcommand{\sio}{SiO $J=2-1$}

\newcommand{\kms}{km s$^{-1}$}
\newcommand{\tex}{$T_{\rm{ex}}$}

\shorttitle{Molecular Line Study of Barnard 1c}
\shortauthors{Matthews, Hogerheijde, J{\o}rgensen \& Bergin}
\begin{document}

\title{The rotating molecular core and precessing outflow of
       the young stellar object Barnard 1c}

\author{Brenda C.~Matthews}
\affil{Herzberg Institute of Astrophysics, National Research Council
  of Canada, 5071 West Saanich Road, Victoria, BC, V9E 2E7, Canada}
\email{brenda.matthews@nrc-cnrc.gc.ca}

\author{Michiel R.~Hogerheijde}
\affil{Leiden Observatory, P.O. Box 9513, 2300 RA, Leiden, The Netherlands}
\email{michiel@strw.leidenuniv.nl}

\author{Jes K.~J{\o}rgensen}
\affil{Harvard-Smithsonian Center for Astrophysics, 60 Garden Street
  MS42, Cambridge, MA, 02138}
\email{jjorgensen@cfa.harvard.edu}

\and

\author{Edwin A.~Bergin} \affil{Department of Astronomy, University of
Michigan, 825 Dennison Building, 500 Church Street, Ann Arbor, MI
48109} \email{ebergin@umich.edu}

\begin{abstract}
We investigate the structure of the core surrounding the recently
identified deeply embedded young stellar object Barnard 1c which has
an unusual polarization pattern as traced in submillimeter dust
emission.  Barnard 1c lies within the Perseus molecular cloud at a
distance of 250 pc.  It is a deeply embedded core of 2.4 $M_\odot$
(Kirk et al.) and a luminosity of $4\pm 2 L_\odot$.  Observations (and
resolutions) of \co\ (9.2\arcsec\ $\times$ 5.9\arcsec), \thirteenco\
and \ceighteeno\ (14.3\arcsec\ $\times$ 6.7\arcsec), \hcop\
(7.6\arcsec\ $\times$ 5.8\arcsec), and \ntwohp\ (5.9\arcsec\ $\times$
4.6\arcsec) were obtained with the Berkeley-Illinois-Maryland
Association array, together with the continuum at 3.3 mm (6.4\arcsec\
$\times$ 4.9\arcsec) and 2.7 mm (9.5\arcsec\ $\times$ 6.3\arcsec).
The field of view of the BIMA array antennas at 3 mm is 2.1\arcmin.
Single-dish measurements of \ntwohp\ and \hcop\ with FCRAO reveal the
larger scale emission in these lines with resolutions of 57.5\arcsec\
and 60.5\arcsec, respectively.  The \CO\ and \HCOP\ emission traces
the outflow extending over the full field of view, which coincides in
detail with the S-shaped jet recently found in Spitzer IRAC
imaging. The \NTWOHP\ emission, which anticorrelates spatially with
the \ceighteeno\ emission, originates from a rotating
envelope with effective radius $\sim 2400$ AU and mass $2.1 - 2.9 \
M_\odot$, as derived from the 3.3 mm continuum emission.  \NTWOHP\ emission 
is absent from a 600~AU diameter region around the young star, offset from the
continuum peak.  The remaining \NTWOHP\ emission may lie in a coherent
torus of dense material. With its outflow and rotating envelope,
B1c closely resembles the previously studied object L483-mm, and
we conclude that it is a protostar in an early stage of evolution,
i.e., Class 0 or in transition between Class 0 and Class I.  We
hypothesize that heating by the outflow and star has desorbed CO from
grains which has destroyed \NTWOHP\ in the inner region and surmise
that the presence of grains without ice mantles in this warm inner
region can explain the unusual polarization signature from B1c.
\end{abstract}


\section{Introduction}

Class 0 sources represent the youngest phase of low-mass star
formation.  They are characterized by higher infall rates than more
evolved sources based on outflow activity \citep{bon96,ww01}, an
absence of optical emission, and a high ratio of submillimeter to
bolometric luminosity.  Observations of the internal structure of
Class 0 objects rely on interferometers because the sources remain
deeply embedded within their parent molecular clouds.  To understand
the kinematics of the collapse process requires observation of
multiple molecular lines because the chemistry within these dense,
cold cores is complex.  Carbon-bearing species are observed to be
strongly depleted within the core interiors due to freeze-out onto
grains. Nitrogen-bearing species were thought to deplete much more
slowly than carbon-bearing species because of the low binding energy
of the N$_2$ molecule, requiring long periods of time or very high
densities to show depletion \citep{ber97,ber02}.  However, recent
laboratory results indicate that CO binding energies are comparable to
those of N$_2$.  These studies also indicate that the degree of mixing
of N$_2$ and CO ices impacts the desorption timescales
\citep{obe05,bis06}.  Nonetheless, it is clear that \NTWOHP\ (and its
daughter product NH$_3$) show a strong rise in abundance as its main
destroyer, \CO, depletes from the gas, making \NTWOHP\ an effective
tracer of internal core kinematics \citep[e.g.,][]{aik05}.  When \CO\
is present however, reactions between C$^+$ or \CO\ lead to the
eventual destruction of the two standard nitrogen-bearing tracers:
NH$_3$ and \NTWOHP.

Recent observations of several Class 0 sources have shown an absence
of \NTWOHP\ emission at the source position on scales within several
hundred AU of the protostar \citep[i.e.,][]{jor04b}.  Typically,
\NTWOHP\ emission is depressed at the core center and has been
interpreted as destruction of \NTWOHP\ due to evaporation of CO and
its isotopes from dust grains \citep[e.g., L483-mm (hereafter
L483),][]{jor04} or depletion of \NTWOHP\ at high densities
\citep[e.g., IRAM 04191+1522 (hereafter IRAM 04191),][]{bel02}.  An
absence of \NTWOHP\ cannot be purely an indicator of age; L483 is
thought to be in transition between Class 0 and Class I \citep{taf00}
while IRAM 04191 is estimated to be among the youngest Class 0 sources
known \citep{and99}.  Interestingly, in both cases, the \NTWOHP\
emission morphology is double peaked around the source center with a
signature indicating rotation, and an anticorrelation is noted between
\NTWOHP\ emission and the emission from tracers dominated by the
outflow (\CO\ and \HCOP). However, in L483, \CEIGHTEENO\ is centrally
peaked on the source indicating that \CO\ has been evaporated from
dust grains at the core center.  The combination of \NTWOHP\ and
\CEIGHTEENO\ morphology could thus discriminate between whether an
absence of \NTWOHP\ is due to depletion onto dust grains or the
destruction of \NTWOHP\ in the presence of \CO.

In this paper, we present observations of the internal structure of
the protostellar core Barnard 1c (B1c) from the
Berkeley-Illinois-Maryland Association (BIMA) array and the Five
Colleges Astronomical Radio Observatory (FCRAO) 14 m telescope.  B1c
was discovered during 850 \micron\ polarimetry mapping in Barnard 1 by
\citet{mw02}.  Recent IRAC imaging from Spitzer reveals that this
source is highly reddened and deeply embedded in the B1 cloud with an
extensive, highly collimated outflow \citep{jor06}. The presence of
central cavities within young protostellar cores is of particular
relevance to B1c because a heated central region could explain why B1c
has a unique signature in polarized emission.  Its polarization
pattern suggests that the polarized intensity rises to the center of
the core \citep{mw02}, rather than flattening out as seen in other
cores. When compared to the total intensity, a flat distribution in
polarized intensity produces a declining ratio toward the peaks of
cores.  All other low-mass starless and star-forming cores observed in
polarized dust emission have a so-called ``polarization hole'' at high
intensities, thought to arise from changes in magnetic field geometry
or dust grain physics \citep{mat05}.  One of the favored explanations
for the polarization holes is that the grains at core centers are
ineffective polarizers (due to changing grain physics).  Heating
removes the outer grain mantles and could increase the polarization
efficiency of the grains within cavities \citep{whi01}.

Barnard 1 is part of the Perseus molecular cloud complex \citep{bc86a}
which is one of the closest star-forming regions to the Sun.  Its
distance is the subject of some debate, with estimates ranging from
200 pc \citep[based on extinction studies,][]{cer90} to 330-350 pc
\citep[based on the Per OB2 association,][]{dez99,bb64,her83}.  It has
been suggested that the complex may be comprised of two clouds at 200
and 300 pc \citep{cer85}, that there is a distance gradient along its
length \citep{sar79}, or that it is comprised of independent clouds at
varying distances \citep{rid05}.  To facilitate comparison with recent
works on the Perseus cloud \citep[i.e.,][]{eno06,jor06}, we adopt a
distance of $250 \pm 50$ pc to the Barnard 1 cloud, as determined from
recent estimates of extinction \citep{cer03} and measurements of
parallax in members of IC 348 \citep{beli02}.

This paper presents high resolution interferometric data from several
molecular species and continuum emission at 2.7 and 3.3 mm.  The main
objective is to determine whether B1c exhibits a central cavity which
could help explain its unique polarization properties.  The
observations and data reduction techniques are described in $\S$
\ref{obs}.  We present the continuum results and derive the core mass
in $\S$ \ref{dustres}.  The molecular line data are presented in $\S$
\ref{moldatares}.  We discuss these data in $\S$ \ref{disc}.  Our
findings are summarized in $\S$ \ref{sum}.

\section{Observations and Data Reduction}
\label{obs}

\subsection{BIMA Interferometric Data}


\begin{deluxetable*}{lcrccc}
\tablecolumns{6} 
\tablewidth{0pc} 
\tablecaption{Observational Summary of Molecular Species}
\tablehead{\colhead{Transition} & \colhead{Frequency} & \colhead{Date} & 
\colhead{Array} & \colhead{Resolution} & \colhead{Sensitivity}  \\
& \colhead{[GHz]} & & \colhead{Configuration} & \colhead{[\kms]} &
\colhead{[\jybeam]} }
\startdata 
\ntwohp & 93.17378 & 19 November 2002 & C & 0.079 & 0.4 \\
& & & & 0.157  & 0.28 \\
& & 27 February 2003 & B & 0.079 & 0.32 \\
& & & & 0.157 & 0.21 \\
\hline
\\
\hcop & 89.18852 & 24 October 2002 & C & 0.082 & 0.4 \\
& & & & 0.164  & 0.27 \\
& & 25 January 2003 & B & 0.082 & 1.3 \\
& & & & 0.164  & 0.9 \\
& & 11 March 2003 & B & 0.082  & 1.1 \\
& & & & 0.164  & 0.75 \\
& & 16 March 2003 & B & 0.082 & 0.6 \\
& & & & 0.164 & 0.4 \\
& & 07 April 2003 & C & 0.082 & 0.38 \\
& & & & 0.164 & 0.26 \\
\hline
\\
\tablenotemark{a} \sio & 85.64046 & 24 October 2002 & C & 0.085 & 0.38 \\
& & 25 January 2003 & B & & 1.4 \\
& & 11 March 2003 & B & & 1.0 \\
& & 16 March 2003 & B & & 0.5 \\
& & 07 April 2003 & C & & 0.37 \\
\hline
\\
\co & 115.21720 & 04 December 2002 & C & 0.127 & 1.4 \\
\hline
\\
\thirteenco & 110.20135 & 20 November 2002 & C & 0.066 & 0.65 \\
& & & & 0.133  & 0.45 \\
\hline
\\
\ceighteeno & 109.78217 & 20 November 2002 & C & 0.067 & 0.67 \\
& & & & 0.133 & 0.48 \\
\enddata 
\tablenotetext{a}{Not detected.}
\label{obsdetails}
\end{deluxetable*}


Observations were made over the period of 2002 October to 2003 April
using the Berkeley-Illinois-Maryland Association (BIMA) interferometer
\citep{wel96} in Hat Creek, CA.  We utilized a single pointing toward
the position $\alpha_{J2000} = 03^{\rm h}33^{\rm m}$17\fs8,
$\delta_{J2000} = +31$\degr09\arcmin33\farcs0 (J2000).  Two
configurations of the ten 6.1 m antennas were used to observe the
lines of \ntwohp\ and \hcop.  The C-array and B-array had projected
baselines between 2-33 k$\lambda$ and 3-74 k$\lambda$, respectively.
The \co\ line and its isotopomers \thirteenco\ and \ceighteeno\ were
observed only in the C-array configuration.  Table \ref{obsdetails}
contains the sensitivities achieved per track.

The \ntwohp\ line was observed utilizing the digital correlator to
record the line in bands of 6.25 and 12.5 MHz width with 256 channels
each, giving resolutions of 0.079 and 0.157 \kms.  The larger
bandwidth window permitted detections over the range of the seven
hyperfine components of \ntwohp.  All 10 antennas were available for
both 93 GHz tracks.

The \hcop\ line was observed in bandwidths of 6.25 and 12.5 MHz with
256 channels each, resulting in resolutions of 0.082 and 0.164 \kms,
respectively.  A window in the upper side band was senstive to \sio\ in
a 12.5 MHz window with 0.085 \kms\ resolution.  Three tracks were
obtained in the B array and two in the C array. All 10 antennas were
available for three tracks, with one antenna offline for a B array
track and two antennas missing from one of the C array 
tracks.

The \co\ line was observed in a band 12.5 MHz wide across 256
channels.  We also observed it in 100 MHz windows to detect high
velocity CO gas. The velocity resolutions were 0.127 \kms\ and 8.127
\kms, respectively.  No CO emission was detected at velocities
exceeding 20 \kms\ from the rest velocity of the source.  As for the
\ntwohp\ dataset, the phase calibrators were 3c84 and 0237+288.
Only 9 of 10 antennas were useable for this track due to phase
incoherence on antenna 9.  We were sensitive to the CN $J=1-0$ line in
the USB during this observation, but none was detected.

The \thirteenco\ and \ceighteeno\ lines were observed with the digital
correlator configured to record the lines in the upper side band with
12.5 MHz bandwidth over 256 channels. One antenna was offline during
this track. Data from a second antenna was flagged due to the same
phase problems.  Spectral line data from a third antenna was removed
due to noisy phases and amplitudes.

Phase and amplitude variations were calibrated by observing the nearby
quasars 3c84 and 0237+288 (when 3c84 reached elevations exceeding
85$^\circ$) approximately every 30 minutes.  The adopted fluxes of
these quasars was epoch dependent and measured against observations of
the planet Uranus when possible.  The calibration was performed using
the MIRIAD (Multichannel Image Reconstruction, Image Analysis and
Display; \citet{sau95}) task MSELFCAL. 
Absolute flux calibration was done using Uranus when observed or
by derived fluxes of the gain calibrators during the same epoch as our
observations (utilizing the catalogue of fluxes at ``plot$\_$swflux''
on the BIMA website\footnote{http://bima.astro.umd.edu}).  Based on
the uncertainty in flux of the calibrator and the relative variations
in the flux of the quasar, we estimate our overall flux calibration is
accurate to the 30\% level.

Subsequent processing of the data, including the combination of data
from different configurations, were done with MIRIAD.  Images were
produced using MIRIAD's CLEAN algorithm and "robust" weighting
(robustness parameter 0-2) of the visibilities to optimize the
signal-to-noise and the spatial resolution.  Resulting noise levels
are 0.15 Jy/beam in 0.16 \kms\ channels for the \ntwohp\ and \hcop\
line emission, 0.4, 0.6 and 1.0 Jy/beam in 0.13 \kms\ channels for the
\thirteenco, \ceighteeno\ and \co\ line emission respectively.  The
rms levels in the continuum maps are 6.8 and 4.0 \mjybeam\ for the
continuum images at 2.7 and 3mm, respectively.  The best
naturally-weighted resolution is obtained for \ntwohp, with a beam of
FWHM of 5.9\arcsec\ $\times$ 4.6\arcsec.  Moderate resolution is
obtained for \hcop\ and \co\ with FWHM of 7.6\arcsec\ $\times$
5.8\arcsec\ and 9.2\arcsec\ $\times$ 5.9\arcsec\ respectively.  Due to
limited $(u,v)$ coverage, the FWHM of \thirteenco\ and \ceighteeno\ data is
14.3\arcsec\ $\times$ 6.7\arcsec.  Integrated-intensity and
velocity-centroid images were obtained from the cleaned spectral-line
cubes using a 1 or 2$\sigma$ clip level.  The resolution of the
continuum images is 9.5\arcsec\ $\times$ 6.3\arcsec\ and 6.4\arcsec\
$\times$ 4.9\arcsec\ for 2.7 and 3.3 mm respectively.

\subsection{FCRAO Data}

To obtain information on large spatial scales, we observed B1c in
\ntwohp\ and \hcop\ emission with the Five Colleges Radio Astronomical
Observatory (FCRAO).  The data were obtained in very good weather.  We
achieved an rms of 0.08 K ($T_A^*$) in 24 minutes on source.  The
\ntwohp\ and \hcop\ data were obtained simultaneously using the array
SEQUOIA.  The beamsize of the FCRAO data is 57.5\arcsec\ at 93.17378
GHz and 60.5\arcsec\ at 89.188523 GHz.  The maps cover an area
of diameter 12.5\arcmin\ centered on the continuum peak of the BIMA
array data.


\begin{figure*}[t!]
\plotone{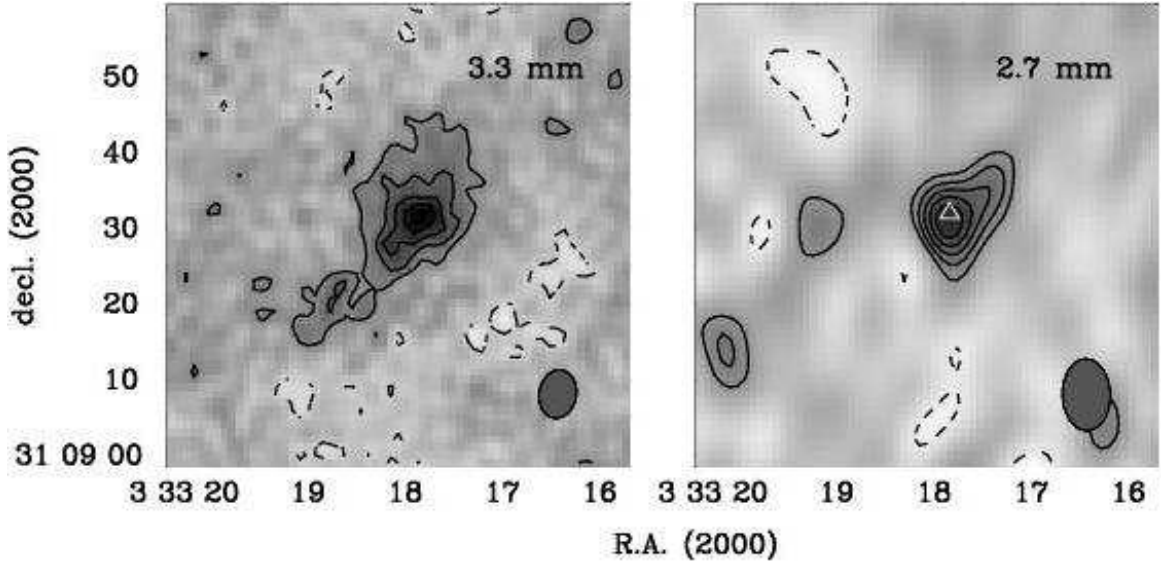}
\caption{Continuum emission from B1c at 3.3 mm
  and 2.7 mm.  The respective beams are shown at the bottom right.
  The 3.3 mm contours are plotted at 2, 4, 6 and 8 $\sigma$ where
  $\sigma = 4.1$ \mjybeam.  The 2.7 mm contours are at levels of 2, 3,
  4, 5, and 6 $\sigma$ where $\sigma = 6.8$ \mjybeam.}
\label{2panel}
\end{figure*}


\subsection{Combination of Single Dish and Interferometer Data}

For the \HCOP\ and \NTWOHP\ data sets, we were able to combine the
BIMA array and FCRAO data to obtain total power maps of each
transition.  We utilized the method described by \citet{sta99} to
combine the data in the image plane.  

The FCRAO data were reordered to match the BIMA axes (x,y,v),
converted to main beam from antenna temperature, and rescaled to
Janskys. The MIRIAD task REGRID was then used to regrid the FCRAO data
to match the parameters of the BIMA array data cube.  Slight
differences in line frequencies were compensated for by a shift in the
reference velocity of the FCRAO spectra prior to regridding.  The
shifts required were 0.885 and 0.016 \kms\ for \ntwohp\ and \hcop\
spectra, respectively.  In the latter case, the shift is significantly
less than the width of individual channels.  A map of the FCRAO beam
for each transition was generated and truncated at the 5\% level,
creating a mask applied to the appropriate single dish data cube.  The
weighting factor for the single dish map was determined by a ratio of
the beam areas.  The composite map was then created using a linear
combination of the BIMA dirty map and the single dish map, followed by
deconvolution using the combined beam \citep{sta99}.

\section{Dust Emission Toward B1c}
\label{dustres}

Figure \ref{2panel} shows the continuum detections toward B1c at 3.3
and 2.7 mm.  The source is located at $\alpha_{J2000} = 03^{\rm
h}33^{\rm m}$17\fs878, $\delta_{J2000} = +31$\degr09\arcmin31\farcs98,
which is the peak of the 3.3 mm emission.  The 2.7 mm emission is
poorly resolved with positive flux density detected to a radius of
13\arcsec.  The 3.3 mm emission is better resolved with the core
extending to 11\arcsec.

The continuum emission is not point-like; Figure \ref{2panel} shows
that it is extended to the northwest and the southeast of the
continuum peak at both wavelengths.  The continuum emission to the
southeast does not coincide at the two wavelengths, but in each case
it is strong, exceeding 4$\sigma$ at 3.3 mm at $\alpha_{J2000} =
03^{\rm h}33^{\rm m}$18\fs75, $\delta_{J2000} =
+31$\degr09\arcmin21\farcs1, corresponding to 16.6 \mjybeam.  The 2.7
mm peak to the southeast also exceeds $3\sigma$ (21.7 \mjybeam) at
$\alpha_{J2000} = 03^{\rm h}33^{\rm m}$20\fs20, $\delta_{J2000} =
+31$\degr09\arcmin14\farcs5.

\begin{figure}
\plotone{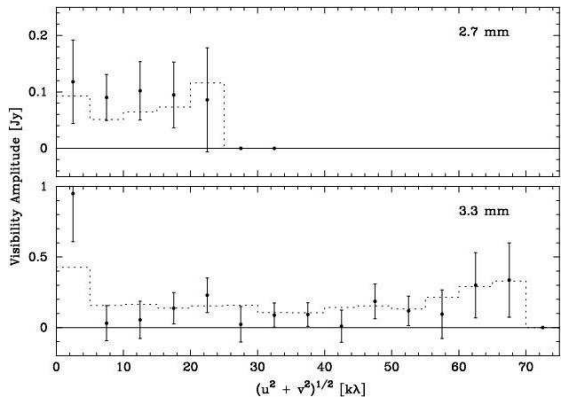}
\caption{Plots of visibility amplitude versus
  projected baseline for our continuum datasets.  Points are the
  amplitudes; the errorbars are the formal standard deviations from
  the mean. The dotted line represents the signal expected from the
  noise.  We have only a single track each at 109 and 115 GHz, both
  in the C array. This is reflected in the extremely limited $(u,v)$
  distance coverage.  Those bins where the detected amplitude exceeds
  the value consistent with noise are even more limited.  The coverage
  is much better at 3.3 mm, where we have multiple tracks at 89 and 93
  GHz, in both the B and C arrays. Similarly, the range of projected
  baselines in which we detect signal well above those consistent with
  noise are also much more extended. } 
\label{ampuvd}
\end{figure}

Figure \ref{ampuvd} shows the visibility amplitude as a function of
$(u,v)$ distance at each wavelength.  Flux detections above the
zero-sigma value (dashed lines) are found for only a few of the
shortest $(u,v)$ distances and are generally marginal detections.
Based on these plots, we interpret the continuum maps as spatially
filtered observations of a resolved, extended envelope, similarly to
the case of L483 (see Figure 2 of \cite{jor04}).  The poorer
$(u,v)$ sampling at 2.7 mm leads to the recovery of less extended
emission than we detect at 3.3 mm.

\begin{deluxetable*}{lcc}[t!]
\tablecolumns{3}
\tablewidth{0pc} 
\tablecaption{Continuum results}
\tablehead{& \colhead{2.7 mm} & \colhead{3.3 mm} }
\startdata
Measured Peak Flux Density [\mjybeam] & $47.5 \pm 14$ & $39.0\pm 13$  \\ 
\tablenotemark{a} Total Flux Density (aperture) [mJy] & $71 \pm 21$ &
$123 \pm 37$ \\ 
Total Flux Density (gaussian fit) [mJy] & $89 \pm 27$ & $150 \pm 50$ \\ 
Size [\arcsec] & $12.5 \times 8.9$ & $13.8 \times 8.8$ \\ 
PA [\degr] & $-51$ & $-30$ \\ 
Deconvolved size [\arcsec] & $10.2 \times 1.6$ & $12.3 \times 7.2$ \\ 
Deconvolved PA [\degr] & $-66$ & $-33$ \\
\hline 
\enddata 
\tablenotetext{a}{Measured in apertures of
13\arcsec\ and 11\arcsec\ radius for 2.7 and 3.3 mm, respectively.}
\label{fluxes}
\end{deluxetable*}


Using the MIRIAD task IMFIT, we have fit gaussians to the 2.7 mm and
3.3 mm maps, fixing the peak flux density and position of the peak
emission.  As Table \ref{fluxes} shows, the resultant total flux
densities derived are higher than those obtained by use of aperture
photometry.  As expected from Figure \ref{ampuvd}, less flux is
recovered at 2.7 mm than at 3.3 mm.  However, the sizes measured for
the core are remarkably similar at each wavelength.  Based on the 3.3
mm deconvolved size, the scale of the detected inner envelope is $\sim
3100 \times 1800$ AU at 250 pc.  The effective radius of this central
region ($r_{\rm eff} = \sqrt{ab}$ where a and b are the major and
minor axes, respectively) is $\sim 2400$ AU.  This is much smaller
than the 12000 AU effective radius derived from the SCUBA map at 850
\micron\ by \citet{kir06}. The measurement from the JCMT incorporates
the entire outer envelope of B1-c, to which our BIMA array data are
not sensitive.  It is normal to measure less flux in interferometric
maps than in single dish data because interferometers preferentially
sample structure on small scales. Therefore, our continuum estimates
only apply to the inner envelope (i.e., not the region traced by the
JCMT).

At 3.3 mm, it is possible that some continuum emission may be due to
free-free emission instead of emission from dust.  We have reduced an
archival observation\footnote{The National Radio Astronomy Observatory
is a facility of the National Science Foundation operated under
cooperative agreement by Associated Universities, Inc..} from the Very
Large Array at 1.3 cm in which B1c lies just outside the primary beam.
There is no detection of 1.3 cm emission in this map, and the $3
\sigma$ upper limit is 3.7 \mjybeam.  Since the free-free emission is
expected to be relatively flat, a contribution of this magnitude
cannot be a significant source of the continuum emission at 3.3 mm.

\subsection{Mass, Column and Density of the Inner Core}
\label{disc-mass}

The mass of the continuum source is easily determined, assuming
optically thin conditions, from the flux of the source and the
temperature via the relation

\begin{equation}
M = \frac{F_\nu \ d^2}{\kappa_\nu \ B_\nu(T_d)}
\label{mass}
\end{equation}

\noindent where $F_\nu$ is the flux, $d$ is the source distance,
$\kappa_\nu$ is the dust opacity and $B_\nu(T_d)$ is the Planck
function at temperature $T_d$.  Table \ref{masses} presents the masses
calculated for our measured flux densities for different assumptions
of $\beta$ (and hence $\kappa_\nu$) and $T_d$.  Our continuum maps are
spatially filtered and are not sensitive to the large scale emission
of the core (i.e., the outer envelope).  If this source is young, then
much of the mass is expected to remain in the envelope; therefore,
observations sensitive only to the inner envelope should be expected
to yields masses smaller than that measured with a single dish
telescope.

\cite{kir06} derive a mass of 2.4 $M_\odot$ for B1c based on SCUBA
observations at 850 \micron.  The clump size ($R_{\rm eff} =
49$\arcsec) and mass were measured with CLUMPFIND \citep{wil94}.  This
estimate of mass and size includes the entire outer envelope, and we
note that the size scale exceeds the maximum scale to which our BIMA
array data are sensitive: 1.1\arcmin\ at 2.7 mm and 1.3\arcmin\ at 3.3
mm.  We concentrate our discussion on the 3.3 mm emission since, as
discussed above, our flux recovery is better at that wavelength. The
requirement that the interferometric mass estimate be less than the
single dish mass estimate allows us to put some constraints on $\beta$
and $T_d$.  The dust temperature is $\ge 15$ K and $\beta$ is more
likely to be closer to 1.0 than to 1.5.  For the same dust temperature
as derived from single dish data (15 K) and $\beta = 1.0$, the mass of
the inner envelope is $2.9 \pm 0.9 M_\odot$.  For $T_d = 20$ K, the
mass is $2.1 \pm 0.6 M_\odot$.  Therefore, the mass of the inner core
lies in the range $2.1 - 2.9 M_\odot$.

Here, we reiterate that our continuum data are hindered by excessive spatial
filtering.  This caveat prevents us from rigorously predicting values of
$\beta$ or dust temperature in the inner core, since the BIMA flux
densities are not easily compared either to each other or the existing
single dish flux density from the JCMT.  

Using the continuum flux, we can also estimate the column density of
molecular hydrogen within the core using the relation:

\begin{equation}
N(H_2) = \frac{S_\nu}{\Omega_m \ \mu \ m_H \ \kappa_\nu \ B_\nu(T_d)}
\label{columnH2}
\end{equation}

\noindent where $S_\nu$ is the peak flux density, $\Omega_m$ is the
main beam solid angle in steradians, $\mu$ is the mean molecular
weight (2.33), $m_H$ is the mass of atomic hydrogen, $\kappa_\nu$ is
the dust opacity per unit mass.  Table \ref{masses} contains estimates
of the column density for different values of $\beta$ and $T_d$.  At
3.3 mm, for a temperature of 20 K, the column density toward the peak
is $(6.8 \pm 2.0) \times 10^{23}$ cm$^{-2}$, for $\beta = 1.0$.
Conversion to extinction using $N({\rm H}_2)/A_V = 10^{21}$ cm$^{2}$ /
mag yields a
visual extinction of $680 \pm 200$ mag to the central peak of the B1c
core.



\begin{deluxetable*}{lcccc}[b!]
\tablecolumns{5} 
\tablewidth{0pc} 
\tablecaption{Masses and Column Densities from Continuum Flux Density}
\tablehead{ & \multicolumn{2}{c}{2.7 mm} &
  \multicolumn{2}{c}{3.3 mm} \\
& \colhead{$\beta = 1$} & \colhead{$\beta = 1.5$} & \colhead{$\beta = 1$} & \colhead{$\beta = 1.5$} }
\startdata 
\tablenotemark{a} $\kappa_\nu$ (cm$^2$ g$^{-1}$) & 0.0049 & 0.0034 & 0.0039 & 0.0025 \\
Mass ($T_d = 12$ K) ($M_\odot$) & $1.2 \pm 0.4$ & $1.73 \pm 0.5$ &
$3.71 \pm 1.1$ & $5.91 \pm 1.8$ \\ 
Mass ($T_d = 15$ K) ($M_\odot$) & $0.92 \pm 0.3$ & $1.32 \pm 0.4$ &
$2.86 \pm 0.9$ & $4.55 \pm 1.4$ \\
Mass ($T_d = 20$ K) ($M_\odot$) & $0.66 \pm 0.2$ & $0.94 \pm 0.3$ &
$2.06 \pm 0.6$ & $3.29 \pm 1.0$ \\
Column Density ($T_d = 12$ K) ($10^{23}$ cm$^{-2}$) & $4.3 \pm 1.3$  &
$6.2 \pm 1.9$ & $12 \pm 4$ & $20 \pm 6$ \\
Column Density ($T_d = 15$ K) ($10^{23}$ cm$^{-2}$) & $3.3 \pm 1.1$ &
$4.7 \pm 1.4$ & $9.5 \pm 2.9$ & $15 \pm 5$ \\ 
Column Density ($T_d = 20$ K) ($10^{23}$ cm$^{-2}$) & $2.4 \pm 0.7$ &
$3.4 \pm 1.0$ & $6.8 \pm 2.0$ & $1.1 \pm 0.3$ \\
\enddata 
\tablenotetext{a}{Derived from $\kappa_\nu = \kappa_0
  (\nu/\nu_0)^\beta$ where $\kappa_0$ is calculated to be 0.01 cm$^2$
  g$^{-1}$ at
  $\nu_0 = 231$ GHz \citep{oss94}.}
\label{masses}
\end{deluxetable*}

\begin{figure*}
\plotone{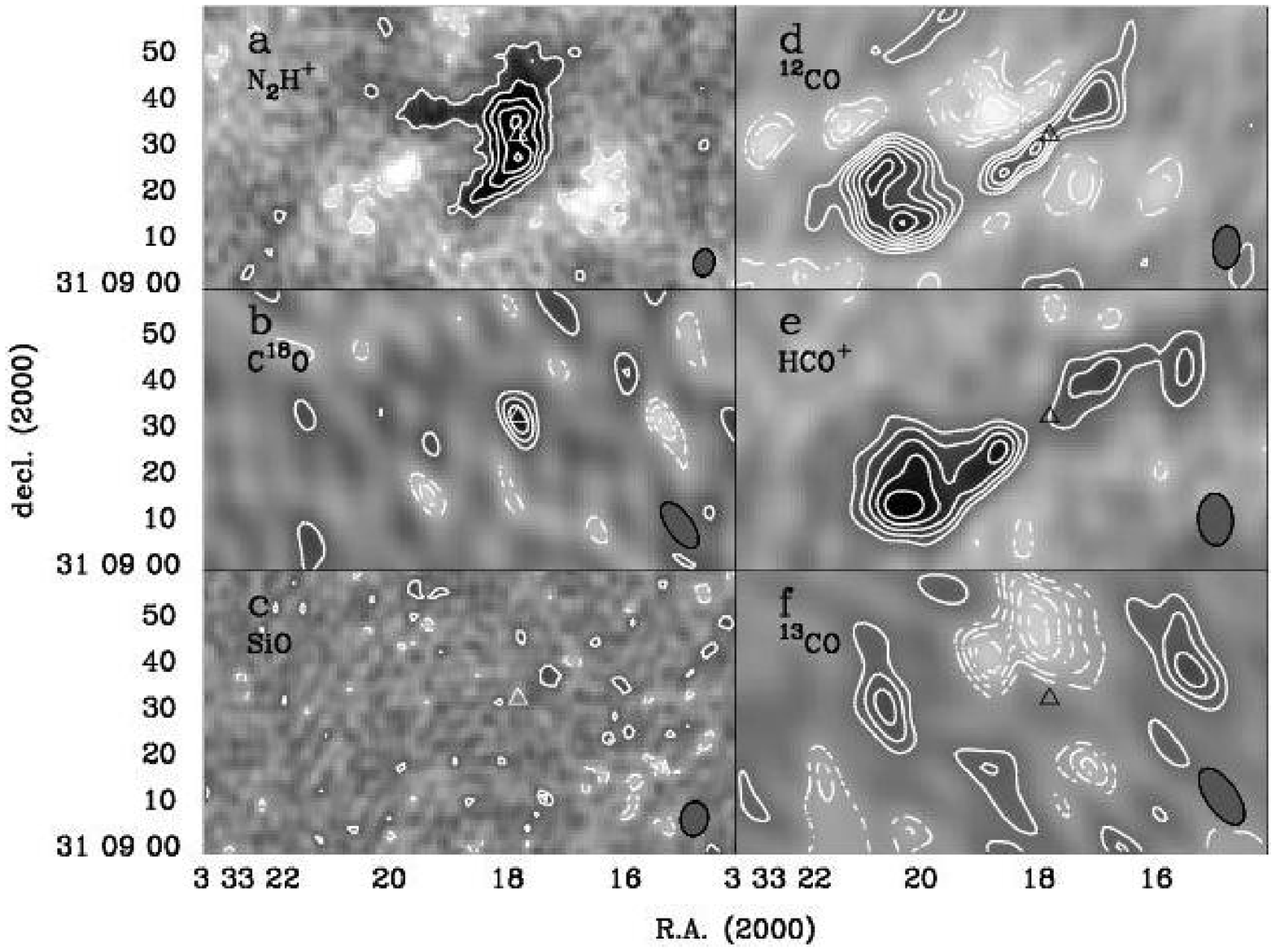}
\caption{Maps of integrated intensity in \co, \thirteenco,
  \ceighteeno, \hcop, \ntwohp\ and \sio\ toward the source Barnard 1c.
  (a) \ntwohp\ map shows the double-peaked distribution with peaks
  offset from the 3.3 mm continuum peak of (c), shown as a
  triangle. The \NTWOHP\ data are integrated over 21.2 \kms\ (a range
  which contains all 7 hyperfine components); contours range from 2 to
  10 $\sigma$ in steps of $2\sigma$ where $\sigma = 47$
  \mjybeam. Dashed contours show the corresponding negative
  contours. (b) \ceighteeno\ data clearly trace the core emission and
  are centrally peaked on the 3.3 mm continuum source peak
  (triangle). The \CEIGHTEENO\ data are uniform weighted for better
  resolution (11.7\arcsec\ $\times$ 5.4\arcsec) and integrated over
  2.1 \kms, and the contours range from 2 to 4 $\sigma$ in steps of 1
  $\sigma$ = 0.17 \jybeam.  (c) \sio\ data shows no significant
  detections.  Contour levels are 2 and 3$\sigma$ where $\sigma=0.03$
  \jybeam.  (d) \co\ integrated intensity in greyscale and white
  contours; contours range from 2 to 8 $\sigma$ where $\sigma = 0.16$
  \jybeam.  Negative contours of the same range are also plotted. The
  \CO\ map is integrated over 12.7 \kms.  (e) \HCOP\ map in greyscale
  and white contours; data are integrated over 17.2 \kms. The contours
  range from 2 to 6 $\sigma$ where $\sigma=35$ \mjybeam.  (f)
  \thirteenco\ integrated emission over 4.3 \kms reveals no emission
  associated with the continuum peak (triangle). Contours range from 2
  to 4 $\sigma$ in steps of 1 $\sigma = 75$ \mjybeam.  As in other
  plots, the dashed contours represent corresponding negative values.}
\label{panel}
\end{figure*}


Assuming the inner envelope is as deep as it is wide, then we can
crudely estimate the central density by adopting the effective
diameter as the depth and assuming a constant density sphere.  Then
the density is just the column density divided by the depth, and the
central column density is $\sim (9.0 \pm 2.6) \times 10^6$ cm$^{-3}$.
A comparable calculation using 2.7 mm data yields the lower estimate
of $(3.3 \pm 1.0) \times 10^6$ cm$^{-3}$.  Reasonable estimates of the
density therefore range from $(3 - 9) \times 10^6$ cm$^{-3}$.


\section{Molecular Emission Toward B1c}
\label{moldatares}

Maps of the integrated intensity of \co, \thirteenco, \ceighteeno,
\hcop, \ntwohp, \sio\ as detected by the BIMA array are presented in
Figure \ref{panel}.  The outflow from B1c is prominently detected in
\co\ and \hcop.  The \ntwohp, \ceighteeno\ clear trace the envelope
around the central source.  The \thirteenco\ detections show no
emission associated with the core, but may trace some compact features
in the outflow.  \sio\ is not detected.  The systematic velocity of
the source is $6.35 \pm 0.02$ \kms, based on a fit to the \ntwohp\ spectrum at
the position of the continuum source.

\subsection{The Core in \NTWOHP and \CEIGHTEENO}
\label{res-n2hp}

\begin{figure}[h!]
\plotone{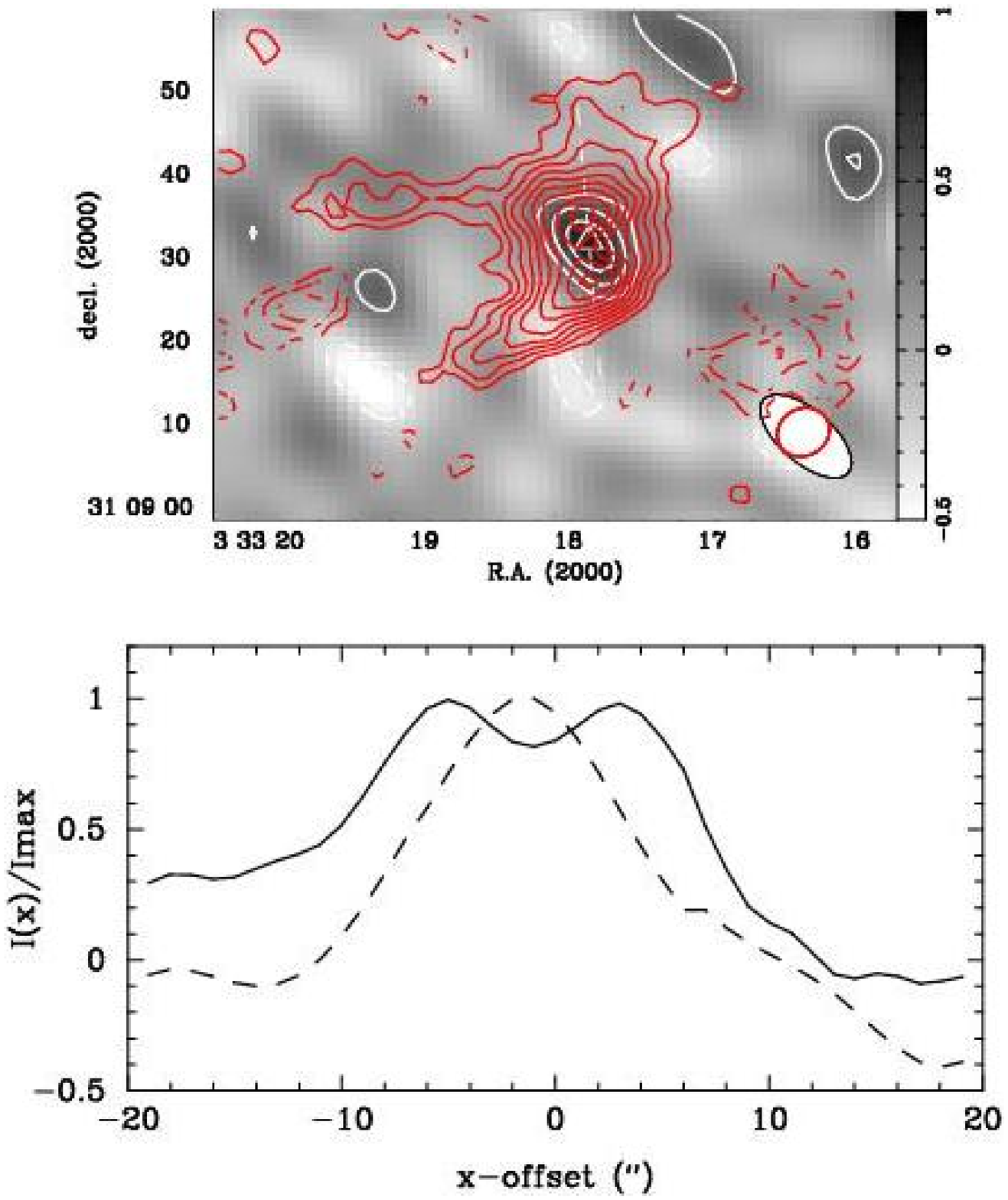}
\caption{A comparison between \NTWOHP\ and
  \CEIGHTEENO\ emission toward B1c.  {\it Upper panel}: The integrated
  \NTWOHP\ (red contours) and \CEIGHTEENO\ (greyscale and white
  contours) emission. The flattened direction of the core is nearly
  north-south and is indicated by the dashed line.  The beams are
  denoted at the lower right. {\it Lower panel}: The \NTWOHP\ (solid
  line) and \CEIGHTEENO\ (dashed line) emission along a slice denoted
  by the dashed line in the upper panel.  The anti-correlation between
  the peaks is evident. The ordinate value is the ratio of the flux
  density at the position along the slice relative to the peak flux
  density in the field.}
\label{n2hp_c18o}
\end{figure}

\ntwohp\ is strongly detected in B1c, indicating dense, cold gas
within the core.  Figure \ref{panel}a shows that the integrated
intensity of \ntwohp\ exhibits a double-peaked structure, with both
peaks avoiding the position of the continuum source (indicated by the
triangle) and the outflow, as shown in Figures \ref{panel}d and
\ref{panel}e.  In contrast, Figure \ref{panel}b shows that the
\ceighteeno\ emission is strongly centrally peaked very close to the
continuum source position.  Figure \ref{n2hp_c18o} shows a comparison
of the distributions of the \CEIGHTEENO\ and \NTWOHP\ emission.  The
emission in a strip along the flattened axis of the core is also
shown.  This figure clearly illustrates the relative positions of the
peaks in the \ceighteeno\ and \ntwohp\ emission.  Despite our
significantly poorer resolution in \ceighteeno, the distribution is
strikingly similar to that observed toward L483 (Figure 10 of
\cite{jor04}).  The relative strength of the emission at each \NTWOHP\
peak is comparable, which differs from the results in L483, in which
one peak was noticably brighter than the other.

\begin{figure}[h!]
\plotone{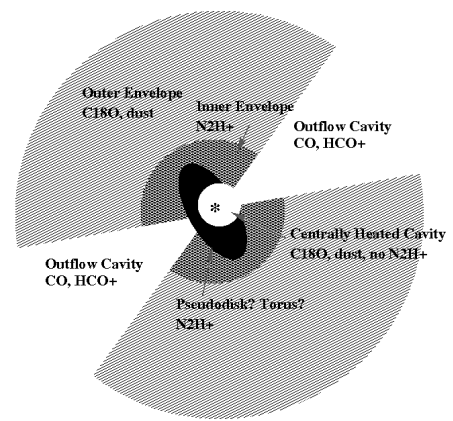}
\caption{A schematic representation
  of the main features of the B1c source.  Our interferometric data
  are insensitive to the large scale outer envelope.  We detect dense
  material from the inner envelope in \NTWOHP\ emission, some of which
  may be confined in a rotating torus centered on the source. The
  outflow, detected in \CO\ and \HCOP\ has carved cavities in the
  inner (and presumably outer) envelope.  A centrally heated cavity is
  indicated by the anticorrelation of \CEIGHTEENO\  and \NTWOHP\ at
  the core center.  Dust emission remains centrally peaked on the
  protostar because very high temperatures (2000 K) are required for
  its destruction.} 
\label{schematic}
\end{figure}

This morphology is seen across a wide range of young protostellar
objects.  It is very similar to that observed by \citet{jor04} in the
more luminous ($9 L_\odot$) evolved Class 0 source L483, and is also
seen in NGC1333 IRAS 2 \citep{jor04a} and sources in the Serpens
molecular cloud \citep{hog05}.  It is also noted in the very low
luminosity object (VeLLO) IRAM 04191 \citep[$0.15 L_\odot$,][]{lee05}.
In B1c, the \NTWOHP\ emission is not confined merely to the two peaks,
but extensions are detected to the east and southeast which bound the
outflow emission in \CO\ and \HCOP.  These extensions could be the
dense edges of a conical cavity carved into the envelope by the
molecular outflow, as depicted in our schematic diagram of the source
(see Figure \ref{schematic}).  The projected opening angle of this
cavity is $\sim 55\degr$, based on the lowest contour of the eastern
extensions of \NTWOHP\ emission in the integrated intensity images
(Figure \ref{panel}a) relative to the position of the continuum source.
The uncertainty on the opening angle of the cavity is $\sim
10\degr$.  A third extension in \NTWOHP\ is observed
to the northwest along the 3$\sigma$ contours of the \HCOP\ and CO
emission.  No \NTWOHP\ is detected along the other edge of the
red-shifted outflow emission in the integrated intensity map.

\begin{figure*}
\plotone{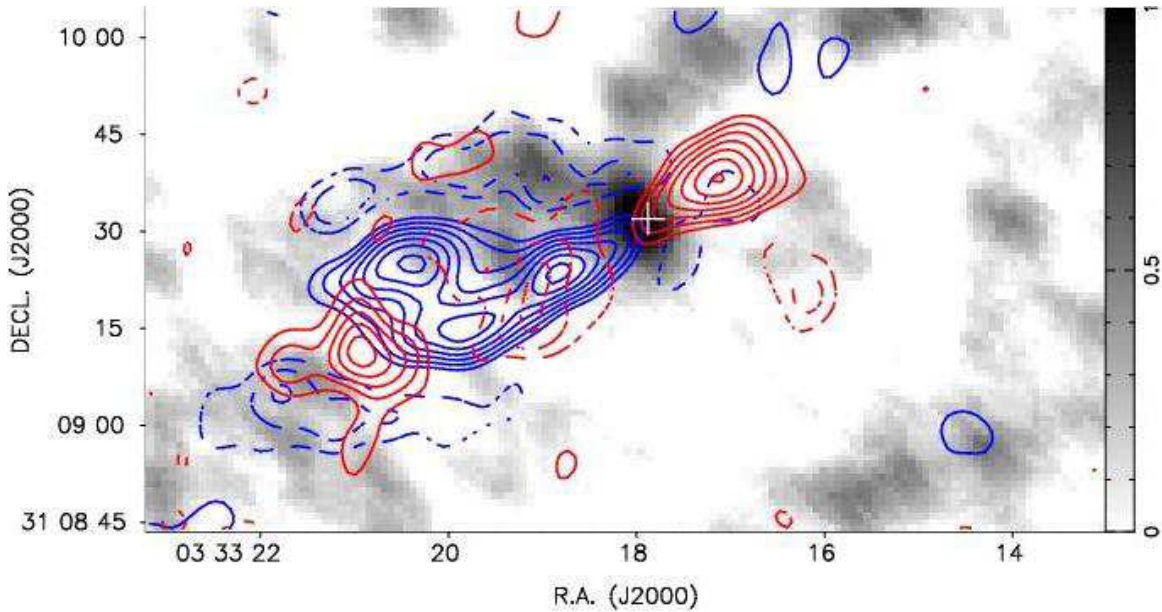}
\caption{The moment 0 data for the red and blue CO
outflow lobes toward B1c.  The greyscale is \ceighteeno\ emission
from the core; the blue contours indicate blue-shifted emission (2.4
-- 5.1 \kms) and the red contours are the red-shifted emission (8.4 --
13.8 \kms). Some red-shifted emission is detected to the southeast of
the blue lobe, indicating that the back of the conical cavity is being
detected, or that the outflow is shifting direction at this location.}
\label{co_moments}
\end{figure*}


\subsection{The Outflow in \HCOP\ and \CO}
\label{res-outflow}

The molecular outflow is clearly seen in both \co\ and \hcop, as shown
in Figure \ref{panel}d and \ref{panel}e.  We did not detect any \sio\
emission as shown by Figure \ref{panel}c.  The detected \co\ and
\hcop\ emission exhibits a bipolar morphology with the blue lobe
emission to the southeast much stronger than the red lobe emission to
the northwest.  Significantly, some \co\ emission is also detected
coincident with the central core, but \hcop\ is not detected toward
the position of the continuum source.  As observed in L483 by
\cite{jor04a}, the \NTWOHP\ data is anticorrelated with the \HCOP\
emission.

The zeroth moment of \CO\ and \CEIGHTEENO\ data are compared in Figure
\ref{co_moments}.  The \ceighteeno\ emission is highly centrally
peaked on the continuum source and is absent along the outflow axis,
due to it relatively low abundance. The \co\ outflow appears to have
carved out the lower density envelope traced by \ceighteeno\
(integrated over the central velocities associated with the source) in
a similar manner as the very dense gas traced by \ntwohp\ in Figure
\ref{panel}a.  In fact, the optically thin species \ceighteeno\ traces
well both the continuum peak and small scale structure within the
inner envelope, detected on the edge of the \co\ emission along both
the blue and red lobes.  This is consistent with the findings of
\cite{arc06} in their survey of outflow sources with the Caltech
Millimeter Array.  The orientation of the outflow from B1c is
approximately $-55$\degr\ east of north.  Based on the blue lobe, the
opening angle appears to be $\sim 35$\degr\ in projection.  Figure
\ref{co_moments} also clearly indicates the presence of red-shifted CO
emission at the leading edge of the blue lobe.


\subsubsection{Spitzer Mid-infrared Emission}
\label{res-jet}

\begin{figure*}[t!]
\plotone{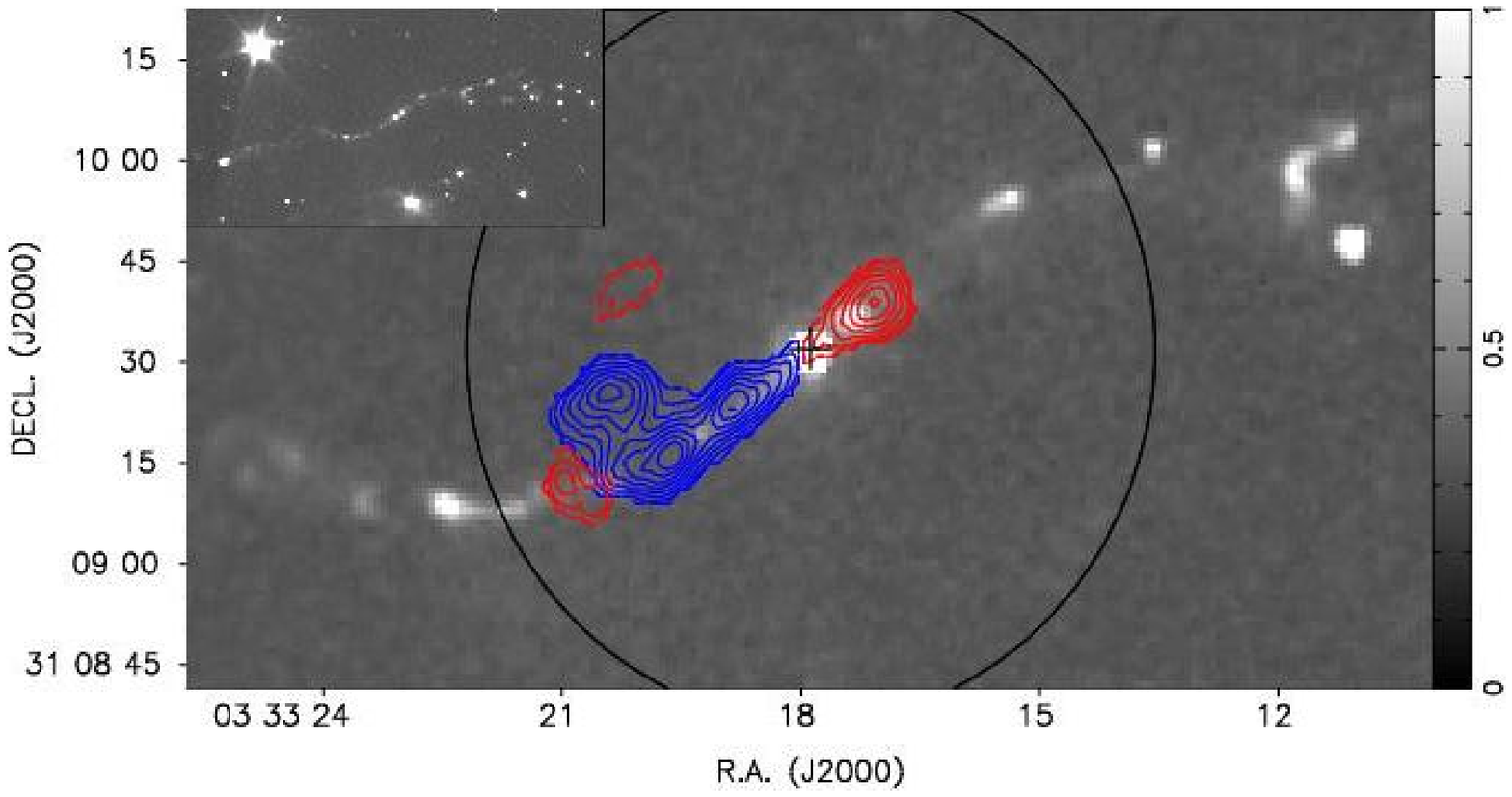}
\caption{IRAC Spitzer data at 4.5 \micron\ from
\citet{jor06} are compared to moment maps of the \co\ emission over
the blue and red lobes.  The correlation between the position angle
and central axis of the outflow is excellent between the near-IR and
millimeter data. The inset shows the larger scale extent of the
molecular hydrogen emission seen from Spitzer. \co\ contours are from
3-10 \jybeam\ \kms, in steps of 1 \jybeam\ \kms.  The cross marks the
continuum peak of the BIMA data.  The large circle indicates primary
beam of a BIMA antenna.}
\label{spitzer_co}
\end{figure*}

Figure \ref{spitzer_co} shows the outflow emission from the source B1c
as imaged by Spitzer with IRAC at 4.5 \micron\ \citep{jor06} and the
\CO\ outflow emission from the BIMA array.  The IRAC 4.5 \micron\ band
has been found to be a strong tracer of outflows which is likely due
to the  presence of H$_2$ pure rotational transitions and the CO
fundamental vibrational mode within the bandpass \citep{nor04}.  The
outflow clearly extends for several arcminutes on either side of the
driving source, B1c.  The 4.5 \micron\ emission is a strong tracer of
molecular hydrogen, typically associated with protostellar jets.  An
``S-shaped'' morphology is evident in the Spitzer map.  Such S-shaped
jets are interpreted as precession of the jet because of the presence
of an (unseen) binary \citep[e.g., ][]{hod05}.  Near the source, the
outflow appears quite symmetric; however, at larger distances, the
blue lobe widens while the red lobe appears to either split into two
separate sequences of ``bullets'' or be confused with an outflow from
a different source.  The projected linear extent of the outflow
detected by Spitzer is dependent on which distance one adopts to the
Barnard 1 cloud.  At 250 pc, the 6\arcmin\ extent of the blue lobe
from the central source indicates a distance of $\sim 90000$ AU, or
0.44 pc.

Comparison of the Spitzer data with the BIMA array data illustrates
that the 4.5 \micron\ emission lies along the central axis of the CO
outflow very near the driving source. Like the molecular hydrogen
emission, there is a bend in the \CO\ emission, located precisely where
the CO emission transitions from blue-shifted to red-shifted emission.
We are constrained by the single pointing of the BIMA observations to
a single primary beam of coverage in the CO emission.  More extended
\co\ measurements are needed to compare the morphology of the
molecular hydrogen (the driving jet) to the morphology of the
entrained gas mapped by the \CO.  It is interesting that the position
of the Spitzer emission peaks are anti-correlated with the CO peaks in
the BIMA data, which is likely due to the difference in excitation
conditions between the H$_2$ transitions that sample warm (few hundred
K) gas along the jet shocks and \co\ which preferentially traces
entrained, cold ($\sim 10$ K) gas.

\subsection{\thirteenco\ Emission}

Figure \ref{panel}f shows the integrated emission detected from
\thirteenco.  In contrast to the other CO isotopologues, the
\THIRTEENCO\ data do not clearly trace the core or the outflow.  This
is likely due to missing short spacings in the interferometric map
which limit the sensitivity of this observation both to structure and
emission.  The peaks detected in \THIRTEENCO\ may be associated with
the outflow or the walls of a cavity carved by the outflow.

\subsection{Moment Maps of the Core and Outflow}
\label{moments}

We produced moment maps of the combined \ntwohp\ emission from the
FCRAO and BIMA array datasets over a velocity range of 1.57 \kms, or 10
channels.  After creating the moment maps, we masked out all data
values in the moment zero and moment one maps at positions with values
less than 0.24 \jybeam\ \kms\ in the moment zero map, which is
approximately 1.5 times the rms level per channel, 0.14 \jybeam.

\begin{figure}
\plotone{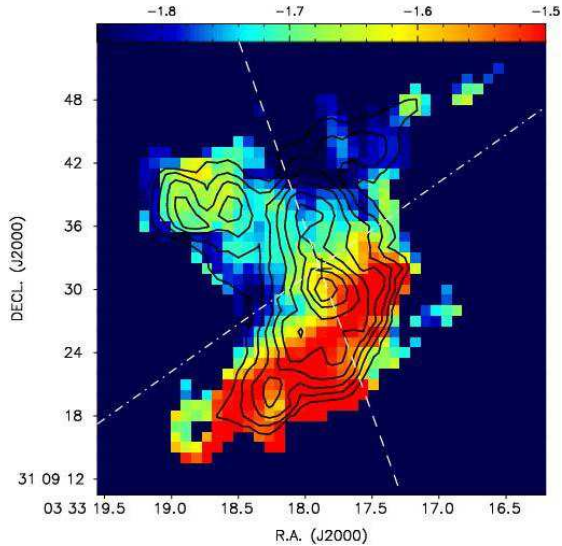}
\caption{First (greyscale) and zeroth (contour) moment maps
from the combined FCRAO and BIMA array observations of \NTWOHP\ 1-0
emission.  The contours are intervals of 0.5$\sigma$ from 1.5 to 6
$\sigma$ where $\sigma = 0.115$ \jybeam\ \kms.  The cross marks the
position of the continuum peak. These maps are taken over the single
isolated component of the \NTWOHP\ hyperfine transitions, F$_1$ F =
0,1 $\rightarrow$ 1,2. The white dashed line represents the slice
taken to produce the position vs.\ velocity diagram of Figure
\ref{pvcuts}, and defines the orientation of the ``torus'' at 10\degr\
east of north. The dot-dashed line represents the position angle of
the outflow (-55\degr\ east of north) as measured from IRAC 4.5
\micron\ and BIMA \co\ data (see $\S$ \ref{res-outflow}).}
\label{comb_n2hp}
\end{figure}


\begin{figure*}[th!]
\plotone{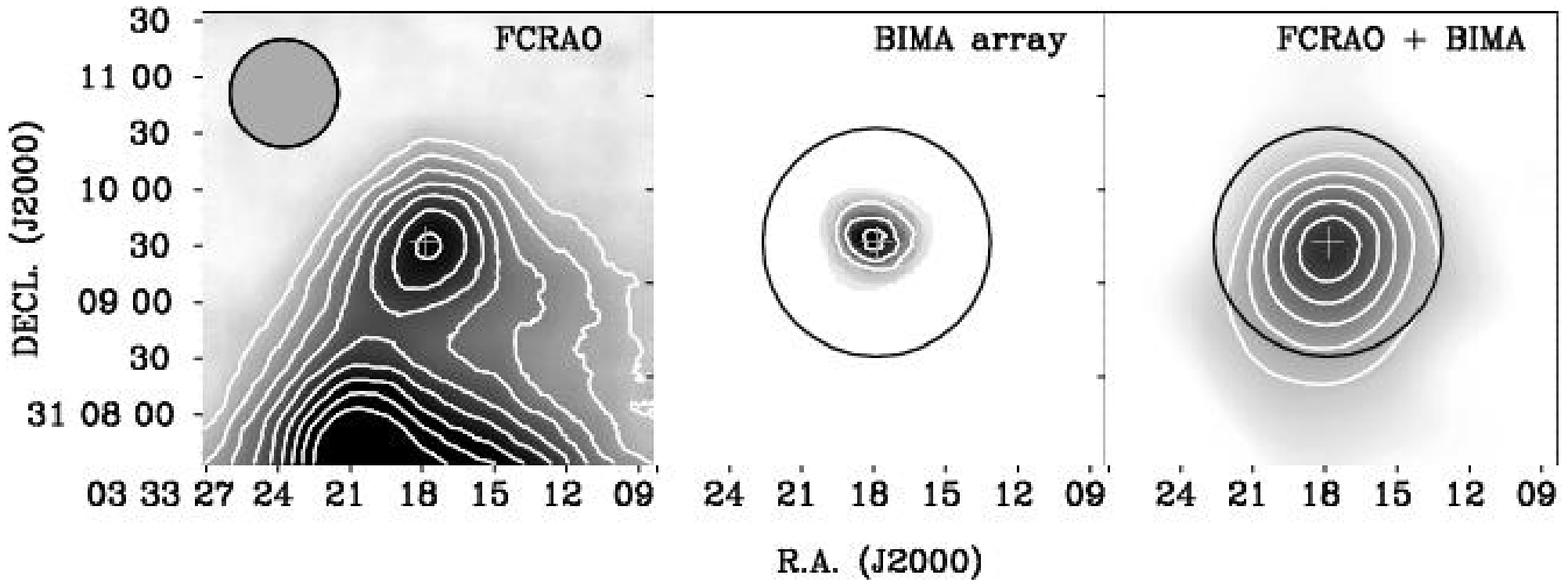}
\caption{Comparison of the zeroth moment
  \ntwohp\ emission on large scales from FCRAO, the BIMA array and
  combined FCRAO and BIMA array data, where the latter two maps have
  been convolved to the FCRAO beam size of 57.5\arcsec, shown at the
  top left of the FCRAO image.  These maps are taken over all seven
  hyperfine components.  The contours of the FCRAO and combined maps
  are at intervals of 10$\sigma$ from 20 to 100 $\sigma$ where $\sigma
  = 1.7$ \jybeam\ \kms.  The BIMA array map has substantially less flux
  than either the FCRAO or combined maps. The contours plotted range
  from 4 to 12 \jybeam\ \kms, in steps of 4 \jybeam\ \kms.  A cross
  marks the position of the 3 mm source.  The BIMA array data,
  convolved to a similar beam, peaks at the same position as the FCRAO
  data alone.}
\label{n2hp_fcrao}
\end{figure*}


Figure \ref{comb_n2hp} shows the first and zeroth moments of the
combined \ntwohp\ emission.  The velocity gradient suggests that the
molecular core is rotating about an axis aligned with the outflow.  As
in the integrated intensity map from the BIMA array of Figure
\ref{panel}a, the zeroth moment map of the combined dataset
illustrates a double-peaked distribution with peaks offset from the
position of the continuum emission.  This morphology is expected from
the projection of a torus of dense gas surrounding the core center
(see Figure \ref{schematic}).  We similarly interpret this
double-peaked feature as a rotating torus with the blue-shifted
emission lying predominantly to the north and the red-shifted emission
lying to the south as indicated by fits to the spectra and the
features of the first moment map. Figure \ref{comb_n2hp} shows the
moment one map in greyscale and the moment zero map in contours.  We
do not have sufficient resolution in our \ntwohp\ spectra to separate
out all the hyperfine splitting (hfs) components in the spectra.  The
presence of the hfs structure complicates the interpretation of the
velocity field from the line emission, which can be alleviated by
using the isolated component.  The moment map is taken over the
isolated line component of \ntwohp\ F$_1$ F 01 - F12 at a velocity
offset by $-8.0064$ \kms\ from the central line \citep{dor03}.  The
velocity range over which the moments are taken is 1.57 \kms\ (10
channels).

The moment map of \NTWOHP\ shows the same extensions to the
east and northwest as the integrated intensity map shown in Figure
\ref{panel}a. Both the eastern extensions detected in \NTWOHP\ are
red-shifted compared to the adjacent gas, with the south eastern one
containing the most red-shifted gas in the core.  The northwestern
extension appears blue-shifted. A faint extension is detected to the
west of the core which could be the southern edge of the cavity
created by the red-shifted lobe of the outflow.

For comparison, in Figure \ref{n2hp_fcrao}, we show the zeroth moment
map of the FCRAO data alone over the whole mapped area. The B1c source
is part of a broader distribution of dense \ntwohp\ emission which
forms a bridge in emission to the sources to the south: B1a (IRAS
03301+3057) and B1b \citep{hir97}. The distribution is singly peaked,
not surprisingly at the position of the brightest peak detected in the
BIMA and combined maps, as shown in Figure \ref{n2hp_fcrao} through images
of the BIMA array data and combined data convolved to the FCRAO beam
size of 57.5\arcsec.  We note that the \NTWOHP\ emission is comparable
in extent to the 850 \micron\ SCUBA continuum scale of the core
\citep[diameter 1.6\arcmin][]{kir06}.  In contrast, BIMA emission
alone is tracing scales $< 1.3$\arcmin.


The axis of rotation of the combined map of Figure \ref{comb_n2hp}
lies at $\sim -45^\circ$ east of north, which is close to the
$-55$\degr\ orientation angle of the outflow (see $\S$
\ref{res-outflow}), indicated by the dot-dashed line.  However, the
axis of rotation is not orthogonal to the plane of the \NTWOHP\ peaks,
which lies at $\sim 10$\degr\ and is indicated in Figure
\ref{comb_n2hp} by a dashed line.  The offset between the plane of a
potential torus (as estimated from the double peaks) and the rotation
axis as derived from the first moment map is 55\degr, not 90\degr, as
might be expected.  Similarly, the offset between the torus and the
outflow axis is apparently 65\degr.

\begin{figure}
\plotone{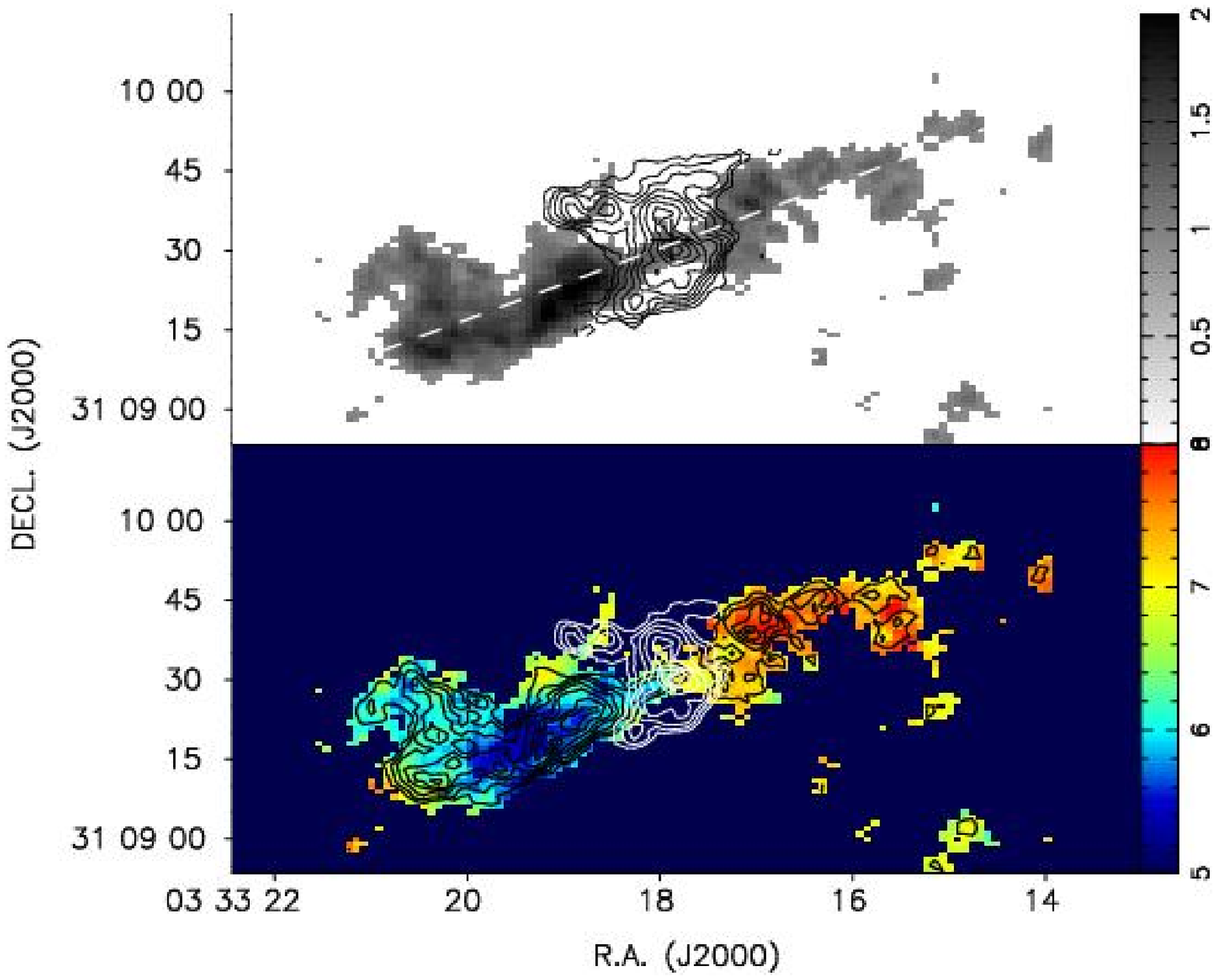}[h!]
\caption{{\it Upper panel:} Moment zero maps of \HCOP\ (greyscale)
and \NTWOHP\ (contours) show the displacement between the \HCOP\ and
\NTWOHP\ emission.  The \HCOP\ emission is confined to a cavity carved
into the dense gas by the outflow.  Contour levels are 2 to 6 $\sigma$
in steps of 0.5 $\sigma$ where $\sigma = 0.11$ \jybeam\ \kms.  The
dashed line shows the position of the position-velocity slice taken
through the outflow (see Figure \ref{pvoutflow}). {\it Lower Panel:} The
first (greyscale) and zeroth (black contours)moment maps of \HCOP\ 1-0
emission from the B1c outflow.  In the combined map of FCRAO and BIMA
data, the blue lobe of the outflow remains brighter than the red lobe.
The white contours show the moment zero \NTWOHP\ emission.}
\label{hcop_n2hp}
\end{figure}

The absence of \NTWOHP\ emission along the outflow axis is likely due
to the carving out of a cavity by the jet associated with the
outflow. The combined FCRAO and BIMA zeroth moment maps of \NTWOHP\
and \HCOP, shown in Figure \ref{hcop_n2hp}, mirror the morphology
observed in the BIMA data alone.  The outflow emission traced by
\hcop\ neatly fills the cavity in the \NTWOHP\ emission.  The most
strongly blue shifted emission of \HCOP\ does not lie at the largest
distances from the central source.  Similar to Figure
\ref{co_moments}, there is evidence for redshifted emission at the
edge of the blue lobe; the outer edge is certainly less blue-shifted
than the material closer to B1c.

\subsection{Kinematics of the Core and Outflow}

\subsubsection{Position-Velocity Diagrams}

\begin{figure}
\plotone{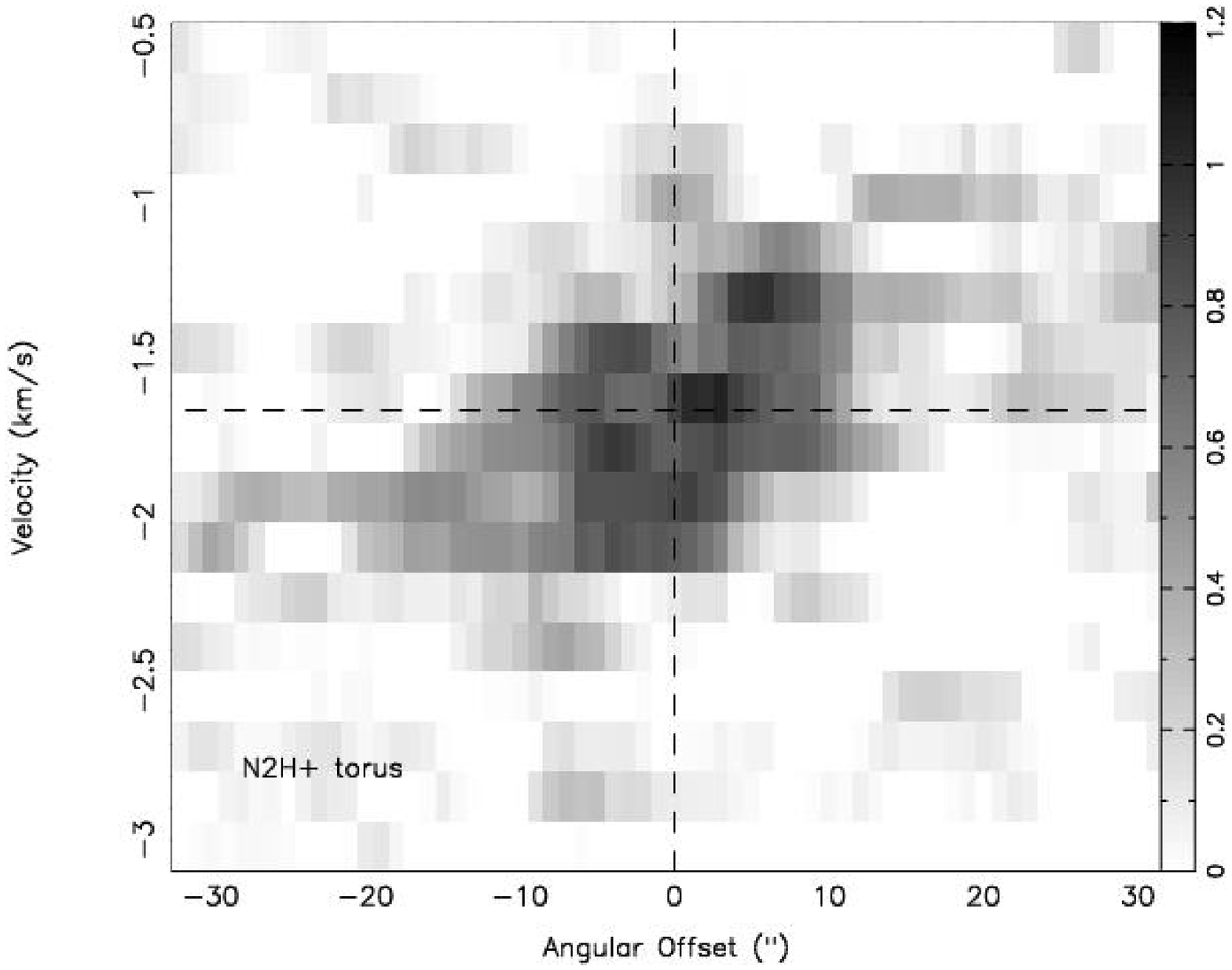}
\caption{Position-velocity diagram of \NTWOHP\ emission
along the plane of the dual peaks observed in the moment map of Figure
\ref{comb_n2hp}.  Only the isolated component of the hyperfine
transitions is used.  The central position is coincident with the
continuum peak. The systemic velocity of the core for the HFS
component is also marked with a dashed line. The velocity gradient
across the core is evident.}
\label{pvcuts}
\end{figure}

We have taken a position-velocity cut along the plane of the torus 
(through the continuum peak position and the two \ntwohp\ peaks) as
shown in Figure \ref{comb_n2hp} using the Karma program ``kpvslice''
\citep{goo96}.  Figure \ref{pvcuts} shows the PV emission of the
isolated \ntwohp\ component.  The emission is continuous across the
molecular core, but emission peaks are well separated in velocity
space.  They are symmetric about the velocity of the source for this
HFS component ($-1.64$ \kms).
The distribution of the emission is very narrow in velocity space,
limited to only a few velocity channels. Nonetheless, there is a clear
velocity gradient across the core with a discernible shift in the
centroid velocity across the two off-center peaks. 

\begin{figure}
\plotone{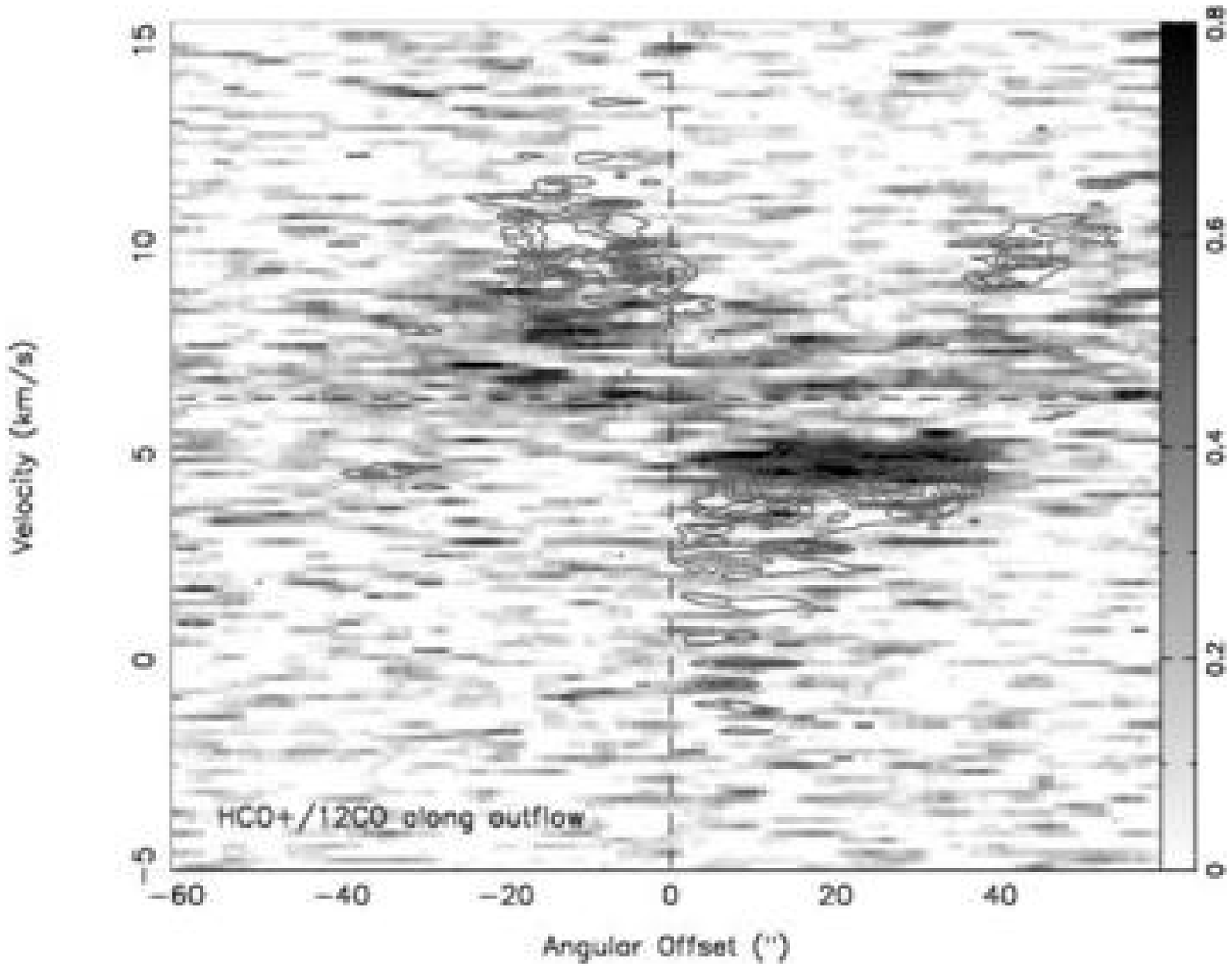}
\caption{The position-velocity diagram of \hcop\ and \co\
emission produced along the outflow axis. The greyscale shows the
\hcop\ emission; \co\ emission (BIMA only) is shown in contours of 20
to 90 percentile in steps of 10 percent.  The slice is centered on the
position of the B1c continuum source. The orientation of the slice is
indicated on Figure \ref{hcop_n2hp}.}
\label{pvoutflow}
\end{figure}

Figure \ref{pvoutflow} shows a position-velocity slice taken along the
outflow axis through the continuum peak. The orientation of the slice
is shown as a dashed line on Figure \ref{hcop_n2hp}.  The distribution
of the outflow in velocity space is relatively confined in the \hcop\
emission (which contains both single dish and interferometric data).
The velocity features are more extensive in the \co\ emission, even
though it is limited to the spatial frequencies detectable to the
interferometer. The red-shifted material seen at the leading edge of
the blue lobe is evident in the PV diagram and no blue-shifted
emission is observed at offsets beyond that boundary.  The red-shifted
material is detected only in the \co\ emission. It is not seen in
\hcop, although Figure \ref{hcop_n2hp} shows that the strongest blue
shifted emission is detected close to B1c. At larger distances from
the source, the \hcop\ emission is dominated by velocities closer to
that of the source. 

\subsubsection{Spectra}

\begin{figure*}
\plotone{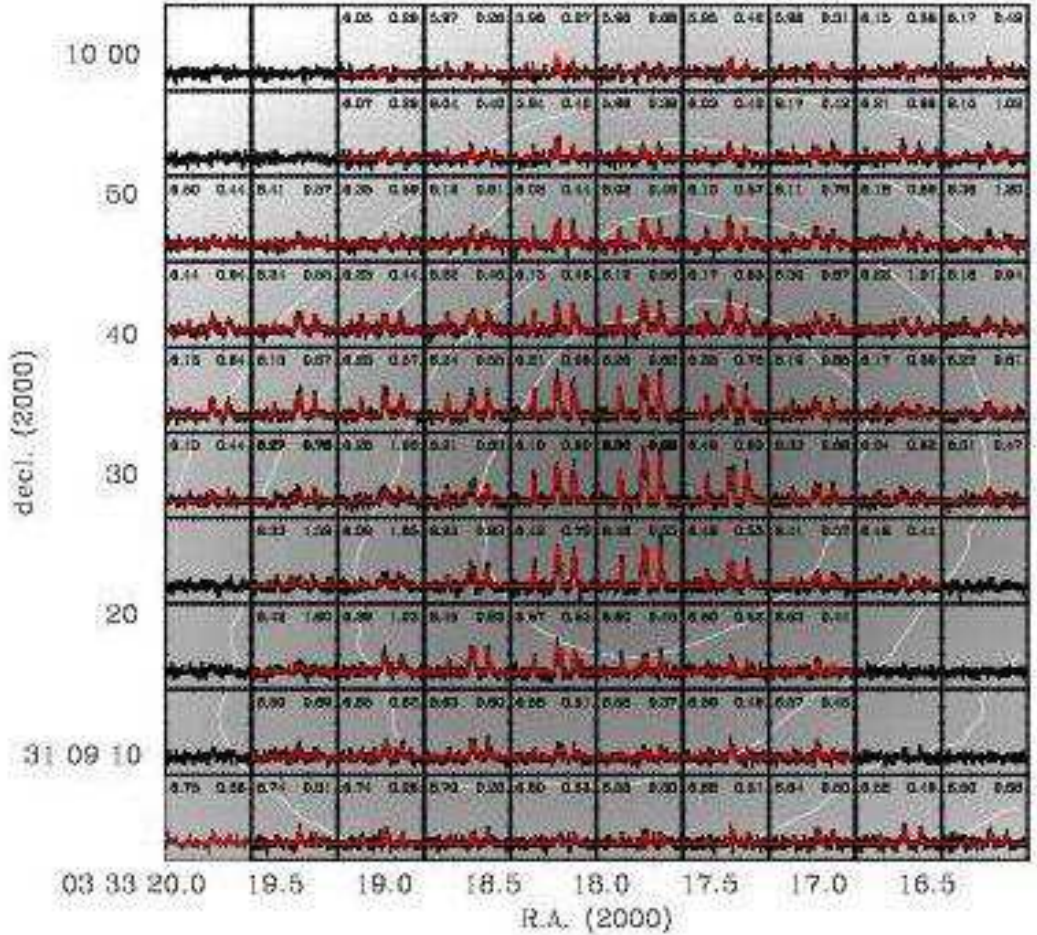}
\caption{\NTWOHP\ spectra across the B1c inner core. The
spectra are integrated across the core in grids of 5\arcsec\ $\times$
6\arcsec, which is slightly greater than the beam area.  Therefore,
the spectra are independently sampling the kinematics of the core.
The velocity range is -10 to 20 \kms, and the intensity range is -0.5
to 2 \jybeam.  Contours of the moment 0 map are plotted below the
spectra.  Contours range from 0.35 to 0.91 \jybeam\ \kms\ in steps of
0.14 \jybeam\ \kms.  The red lines show the fits to the spectra from
CLASS and the fitted values for $V_{LSR}$ and $\Delta V$ are shown at
the left and right corners, respectively, of each grid square.}
\label{spectra}
\end{figure*}

Figure \ref{spectra} shows the distribution of spectra (spatially
binned to an area comparable to the beam size) across the core in the
combined FCRAO and BIMA data.  Fits to the hyperfine components of the
\NTWOHP\ spectra were done using the CLASS software's ``hfs'' fitting
routine \citep{bui02} with up-to-date weights and frequencies
\citep{dor03}.  These fits were used to determine LSR velocities
($V_{LSR}$), intrinsic line widths ($\Delta v$), total optical depths
($\tau(tot)$), and excitation temperatures (\tex).  The distribution
of line strengths is very atypical for \NTWOHP.  The hyperfine ratio
suggests that the central velocity components should be strongest when
compared to higher and lower clusters of hyperfine components.  In
B1c, the central component is often the same strength as the higher
velocity component, which is indicative of high opacity or
self-absorption of the brightest component. Solutions to the hfs
fitting were not substantially improved by using only the outer
triplet and the isolated component of the seven hyperfine lines to
produce fits to the data for the LSR velocity and the linewidth;
therefore, the fits shown in Figure \ref{spectra} are all for the fits
derived with all seven components.

All seven hyperfine components from the combined data were fit
simultaneously in CLASS.  We spatially binned the data onto a grid with spacing
5\arcsec\ in R.A., and 6\arcsec\ in declination.  The fits reflect
centroid velocities ranging from 5.94 to 6.68 \kms.  The range of
realistic line widths (FWHM) is $\sim 0.3$ to 1.5 \kms.  The source
velocities are systematically bluer to the north and redder to the
south as has been discussed in $\S$ \ref{moments}.

\begin{figure*}
\plotone{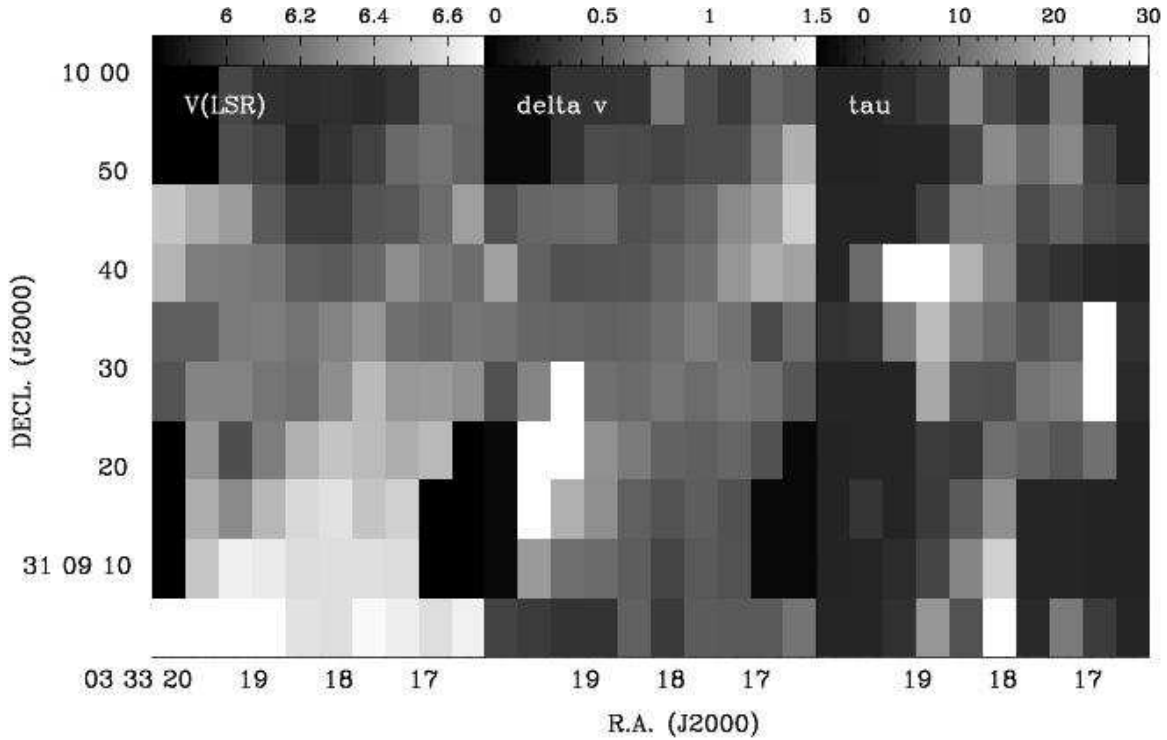}
\caption{Distribution of derived values of $V_{LSR}$, linewidth and
total optical depth based on the fits shown in Figure
\ref{spectra}. The gradient across the source is obvious in the
derived values of $V_{LSR}$.  The linewidths are broader along the
direction of the outflow and are narrower along the plane of the torus
detected in integrated intensity. The optical depths show a large
range of values from optically thin to very optically thick ($>>$ 1),
indicating that several hfs components may be optically thick.  Table
3 shows the mean values of the parameters derived from the fits to
Figure \ref{spectra}.}
\label{deltav_grid}
\end{figure*}

Figure \ref{deltav_grid} shows the distribution of $V_{LSR}$, $\Delta
V$ and line opacity (sum of the peak optical depths for all seven
hyperfine components) across the core on the same grid as shown in
Figure \ref{spectra}.  The gradient in $V_{LSR}$ is well defined
across the field.  Higher values of $\Delta V$ are observed along the
outflow axis where the cavity has been carved in \NTWOHP\ emission
than across the central core where linewidths are quite uniform.  Some
of the best fit solutions (assuming a single excitation temperature
for all components) shown in Figure \ref{spectra} produce very large
total optical depths for the northern and southern emission peaks.
These values indicate that at least one component is very optically
thick and all components may be optically thick.  Attempts to fit the
spectra with lower, fixed estimates of $\tau(tot)$ do an increasingly
poor job of fitting the heights of the components.  The software's
minimum measurable optical depth is 0.1; in these instances, the
emission must be optically thin, but we cannot discriminate between
values of $\tau(tot)$ lower than 0.1.  The maximum value yielded is
30, obtained when components have equal strength. 

\begin{deluxetable}{lrrrr}
\tablecolumns{5} 
\tablewidth{7cm} 
\tablecaption{Parameters Derived from \NTWOHP\ hfs fitting}
\tablehead{ & \colhead{Mean} & \colhead{rms} & \colhead{Min} & \colhead{Max}}
\startdata 
$V_{lsr}$ [\kms] & 6.31 & 2.34 & 5.94 & 6.75 \\
$\Delta V$ [\kms] & 0.61 & 0.37 & 0.26 & 1.85 \\
\tablenotemark{a} $\tau(tot)$  & 6.99 & 8.37 & 0.100 & 30.0 \\
\enddata 
\tablenotetext{a}{The maximum and minimum solutions for $\tau$(tot) based on
  the hfs routine are 0.1 and
  30 respectively.  Values of 0.1 indicate all components are
  optically thin, while 30 indicates that the component strengths are equal.}  
\label{meanfitparams}
\end{deluxetable}

\begin{figure*}
\plotone{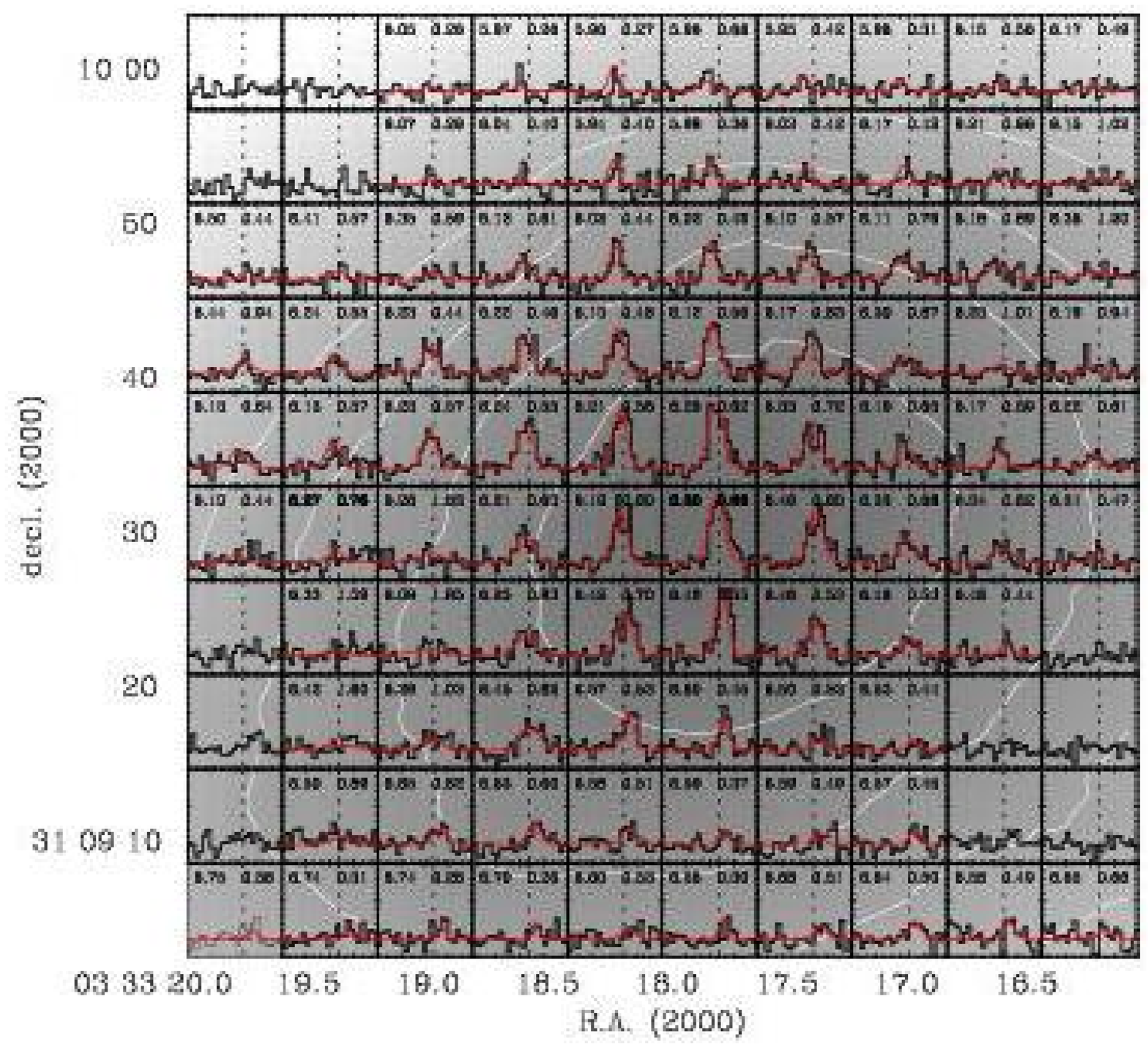}
\caption{\NTWOHP\ spectra of the isolated hyperfine
component of \NTWOHP\ across the B1c core. The spectra are integrated
across the core in grids of 5\arcsec\ $\times$ 6\arcsec, which is
slightly greater than BW sampling.  The velocity range is -4 to 0
\kms, and the intensity range is -0.25 to 1 \jybeam.  Contours of the
moment 0 map are plotted below the spectra.  Contours range from 0.35
to 0.91 \jybeam\ \kms\ in steps of 0.14 \jybeam\ \kms.  The fits shown
as red lines are derived using CLASS for all seven hyperfine
components (Figure \ref{spectra}).  The fitted values for $V_{LSR}$
and $\Delta V$ are shown at the left and right corners, respectively,
of each grid square.}
\label{spectra_isolated}
\end{figure*}

Figure \ref{spectra_isolated} shows only the isolated component of the
\NTWOHP\ spectra, with a velocity range limited to -4 to 0 \kms.  These
spectra reveal differences both in the peak of the lines and in line
shape across the core, with some positions exhibiting a dip in the
center of the line profile, which could indicate that even this line
may be optically thick.  It is also obvious that regions of poorer
S/N produce poorer fits, higher (likely unphysical) estimates of
$\tau$.



\section{Discussion}
\label{disc}

\subsection{Column Density and Chemistry}

\subsubsection{Column Density of \NTWOHP}
\label{n2hpmass}

\begin{figure}
\plotone{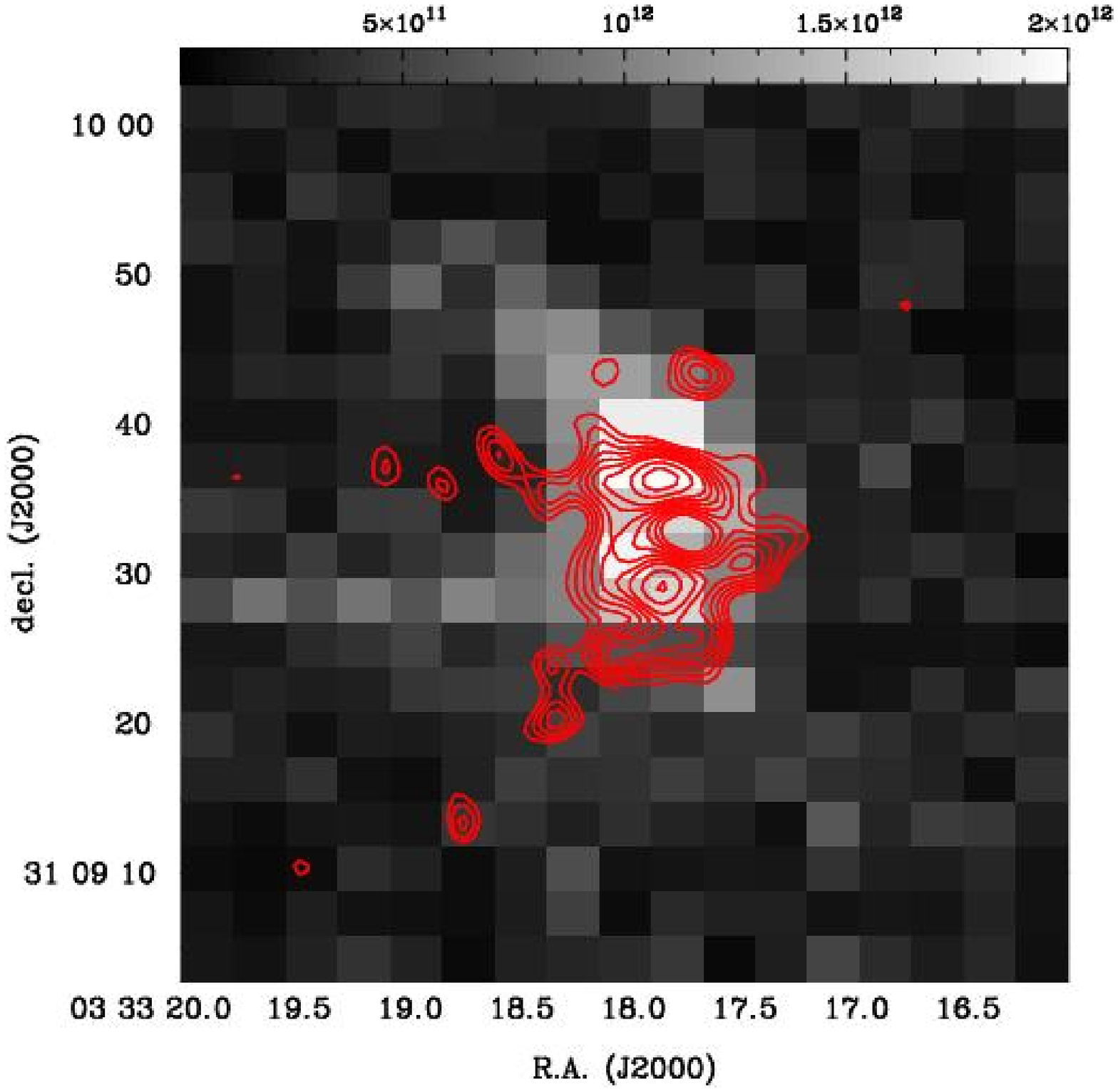}
\caption{Column density distribution of \ntwohp\ emission in 
B1c.  Contours are sum of the \ntwohp\ emission over all channels with
signal exceeding 3 times  the rms level. }
\label{n2hpcolumn}
\end{figure}

In order to estimate the \NTWOHP\ column density, we require estimates
of the FWHM of the lines and estimates of the line opacity. Figure
\ref{deltav_grid} shows the distribution of line widths and opacities
derived from the \ntwohp\ spectra. We derived the \NTWOHP\ column
density using a simple curve of growth determined by assuming LTE
excitation conditions at \tex\ $ = 12$ K, a line width of 0.75 \kms,
and the RADEX escape probability
code\footnote{www.strw.leidenuniv.nl$/ \sim$moldata}
\citep{sho05}. This simple approach takes into account line opacity,
but a more detailed envelope model with proper density and temperature
slope will be required to get more accurate estimates (Matthews et
al. in preparation).  Figure \ref{n2hpcolumn} shows the column density
distribution of \ntwohp\ emission in B1c.  Since the LTE analysis
predicts opacities on the order of unity while the spectra and hfs
component fitting suggest significantly higher opacities, it is likely
that the excitation is not in LTE (i.e., \tex\ $< 12$ K).  
However, Figure \ref{n2hpcolumn} is illustrative as it shows the same
depression in \NTWOHP\ emission seen in the zeroth moment map of 
Figure \ref{uniform}. 

\subsubsection{Column Density of \CEIGHTEENO}
\label{disc-c18o}

To derive the density at the core center, we estimate the column
density at the peak of \ceighteeno\ emission.  We can derive the
column density from the expression
\begin{eqnarray}
\lefteqn{N_{\rm C^{18}O} = \frac{3.34 \times 10^{14}}{\nu \ \mu^2 \ (1 - \exp(-
  h\nu/kT_{\rm ex}))} } \nonumber \\
&& {} \times \ \frac{\int{T_{\rm MB} dV}}{\exp(-Jh\nu/2kT_{\rm
  ex})} \ \frac{\tau}{1 - \exp(-\tau)}
\label{c18ocolumn}
\end{eqnarray}

\noindent where $\nu$ is the frequency in GHz (109.78217 GHz), $\mu$
is the dipole moment in Debye (0.11 Db), \tex\ is the excitation
temperature, and $J$ is the lower rotational level for the transition.
We assume $T_{\rm kin} =$ \tex\ which is constant for all rotational
levels, LTE is valid for the \CEIGHTEENO\ gas and that the optical
depth effects are completely accounted for by the factor $\tau / (1 -
e^{- \tau})$.  We evaluate the integral over the line width from the
peak of the moment zero map, integrated over a linewidth of 1 \kms. We
estimate $\tau$ from the peak temperature of the line via the
expression
\begin{equation}
T_{\rm peak} \approx \frac{h\nu}{k} \ \frac{1 -
  \exp(-\tau)}{\exp(h\nu/kT_{\rm kin} - 1)}
\label{tpeak}
\end{equation}

\noindent where $T_{\rm kin}$ is the kinetic temperature. From the
spectrum at the peak position, we evaluate $\tau$ for $T_{\rm peak} =
2$ K and $T_{\rm kin} \approx T_{\rm ex} = 12$ K.  We find $\tau =
0.244$, which is consistent with the \CEIGHTEENO\ emission being
optically thin.  The strength of the emission at the peak is 1.16
\jybeam\ \kms.  Using the conversion factor for our BIMA array
synthesized beam at 109
GHz (1.12 K/\jybeam), $\int T_{\rm MB} dV$ is 1.3 K \kms.  Substitution of
these values in equation (\ref{c18ocolumn}) yields a column density of
$1.03 \times 10^{15}$ cm$^{-2}$ (the uncertainty, based largely on the
flux calibration, is $\sim 30$\%).  Utilizing the abundance ratio
[C$^{18}$O]/[H$_2$] $= 1.7 \times 10^{-7}$ \citep{fre82} yields an H$_2$
column density of $1.75 \times 10^{22}$ cm$^{-2}$.  The effective
radius of the beam in the \CEIGHTEENO\ observation is $1.8 \times
10^{16}$ cm.  Assuming the core is as deep as it is wide, the density
at the peak is $\sim 9.5 \times 10^5$ cm$^{-3}$.  This value is three times the
estimated central density of $3.2 \times 10^5$ cm$^{-3}$ from the
analysis of dust emission maps by \cite{kir06}, but a magnitude less
than the estimate based on 3.3 mm continuum data in $\S$ \ref{disc-mass}. 
The difference is partly attributable to the different spatial filtering
between the observations. The 2.7 mm data, with more comparable
$(u,v)$ coverage to the \CEIGHTEENO\ data, yields densities of $(3.3 \pm 1.0)
\times 10^6$ cm$^{-3}$, closer to the value derived above.

Aside from the uncertainties introduced by the different degrees of
spatial filtering, the discrepency between the H$_2$ column density
measured with dust and \CEIGHTEENO\ most likely arises due to
depletion from standard ISM abundances, leading to underestimates of
$N({\rm H}_2)$.  Although the grains must be heated in B1c due to the
centrally peaked distribution of \CEIGHTEENO, some of the gas could
remain frozen onto the dust grains.  Assuming the dust represents the
total column of the inner envelope and the gas is depleted, then the
[C$^{18}$O]/[H$_2$] ratio is not $1.7 \times 10^{-7}$ but instead
$1.3 \times 10^{-8}$ (from 2.7 mm dust continuum), a factor of 13
less.  This value is similar to the values measured in Class 0
sources by \cite{jor02}. 

\subsection{A Depression in \NTWOHP\ Emission near the Core Center}

\begin{figure}
\plotone{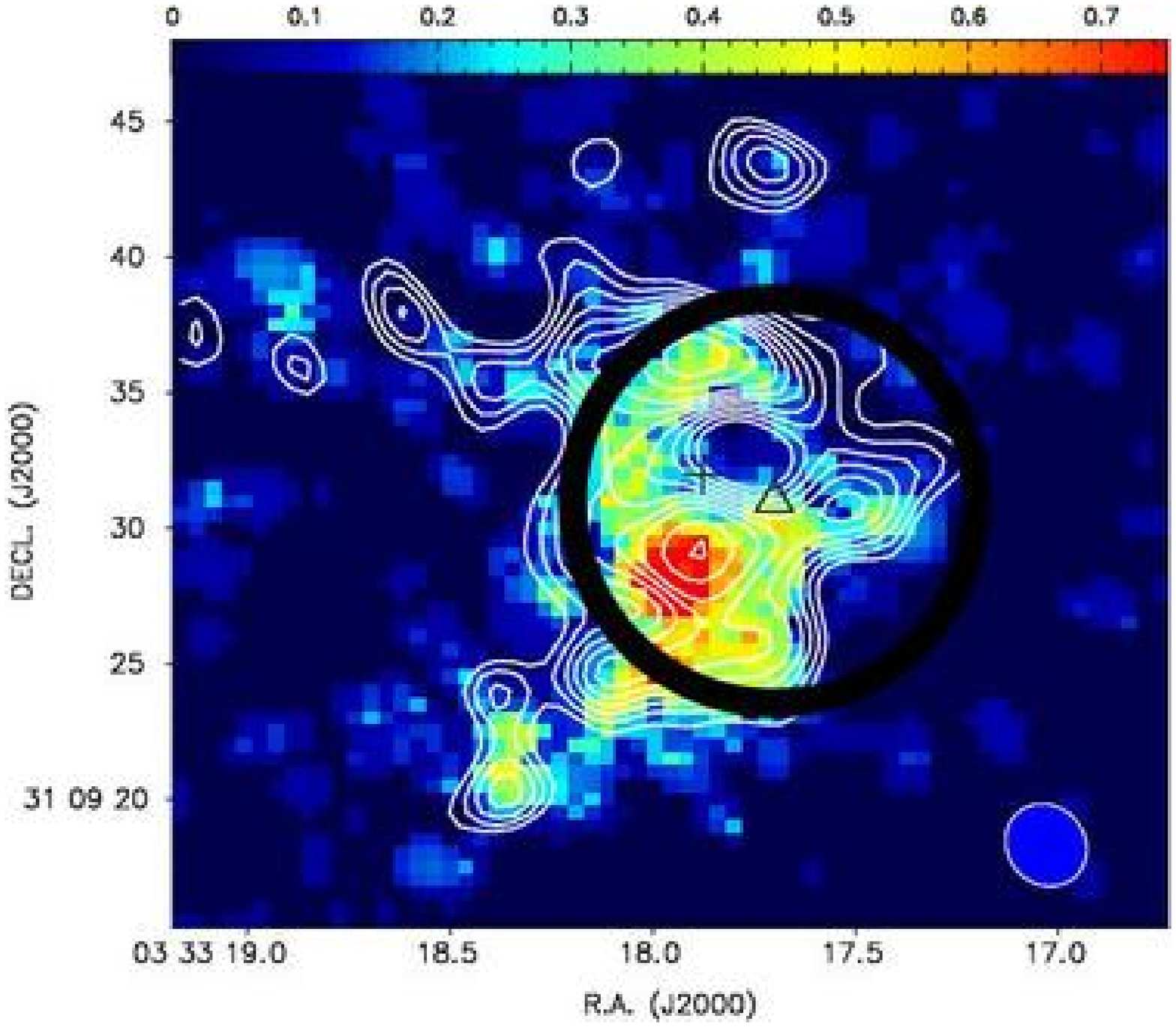}
\caption{Moment zero map of the \NTWOHP\ emission from B1c
with uniform weighting.  The new resolution is 3.09\arcsec\ $\times$
2.52\arcsec, almost a factor of 2 improved over the data in $\S$
\ref{moments}. The cavity can now be seen as an absence of \NTWOHP\
emission offset to the west of the continuum peak, marked by a cross.
The triangle marks the position of the source as determined from the
SCUBA data; the ring depicts the JCMT beam of 14\arcsec.  Contours are
the same as in Figure \ref{n2hpcolumn}.}
\label{uniform}
\end{figure}

As demonstrated by the integrated intensity and moment maps, the
\NTWOHP\ emission is diminished near the core's center.  We are able
to produce a more coherent picture of the central ``cavity'' in
\NTWOHP\ emission by reweighting the \NTWOHP\ data presented in Figure
\ref{comb_n2hp} to uniform when creating the maps. The downweighting
of shorter baselines degrades the S/N of the individual channels
significantly, but the moment map produced retains high signal by
utilizing all seven hyperfine components. Figure \ref{uniform} shows
the moment zero map produced from the uniformly weighted BIMA array
data combined with FCRAO data. A high threshold (3$\sigma$) was placed
on inclusion of individual channels into the moment map to avoid
channels between components.  The cross marks the position of the
continuum source, and the triangle represents the pointing center of
the JCMT observations presented in \citet{mw02}.  The ring depicts the
14\arcsec\ beam of the JCMT.  The scale of the evaporated cavity is
such that it would not have been resolved by the JCMT.  The scale of
the cavity is 2.7\arcsec\ $\times$ 2.15\arcsec, at a position angle of
7\degr.  This corresponds to 675 $\times$ 540 AU at a distance of 250
pc.

The depression of \NTWOHP\ emission in the inner envelope and along the
outflow path is likely due to the destruction of \ \NTWOHP\ by
molecules that are released from grains, most notably CO and H$_2$O. 
Chemical models of an evolving core with a central luminosity source
(i.e., Class 0) suggest that by 10$^5$ years a cavity of $\sim 700$ AU
will be seen in the \ \NTWOHP\ abundance and emission \citep{lbe04}. 
This is the direct result of stellar heating providing enough energy
to evaporate CO from grains within the core center.  This model may
not be directly applicable to B1c due to differences in masses and/or
luminosity, but it provides a rough estimate of the \ time that has
passed since stellar ignition.  In addition, observations and models
suggest that  N$_2$H$^+$ will also be destroyed within the outflow
\citep{bac01,bnm98}, and we see a strong anticorrelation between the
position of the outflow and the presence of \NTWOHP\ emission.

A central depression in \NTWOHP\ emission has also been detected in
the transition Class 0/I source L483 by \citet{jor04} and in the low
luminosity object IRAM 04191 by \citet{bel04}.  \citet{lee05} estimate
the inner edge of the IRAM 04191 cavity has a mean radius of 1400 AU.
This is a factor of four times larger than the cavity detected in B1c.
In a simple model where cavities grow over time, this suggests that B1c could be even
younger than IRAM 04191's estimated age of $3 \times 10^4$ yr since
the onset of accretion \citep{and99}.  The gradients and line widths
measured in \ntwohp\ are strikingly similar in these two cores,
indicating strong dynamical similarities, although we stress that IRAM
04191 is one of the new Class of very low luminosity objects
\citep[VeLLOs,][]{dif06,you04} while B1c is a relatively
luminous ($4 \pm 2 \ L_\odot$) Class 0 object (Matthews et al.\ 2006,
in prep).

We can estimate a kinematic age for this source based on the outflow
data.  The projected linear extent of the emission coupled with the
distance from the source yields estimates of the source age.  Using
the brightest peak of the blue lobe detected in \co\ and assuming an
inclination angle of 45$^\circ$, the dynamical age is $3.7 \times
10^4$ yr, comparable to that of the source IRAM 04191 in Taurus ($3
\times 10^4$ years).  To produce an age on the order of $10^5$ years,
the outflow inclination angle must be $\sim 15$\degr\ (i.e., the
outflow must lie almost along the line of sight).

The absence of \NTWOHP\ in IRAM 04191 is attributed to
depletion within the cold core interior \citep{bel02}.  The strong
centrally peaked detection of \ceighteeno\ emission from B1c (see
Figure \ref{co_moments}) shows that depletion is not significant in
B1c, and the anticorrelation between \NTWOHP\ with the core center and
outflow lobes suggests that \NTWOHP\ is absent due to destruction by
CO.  In addition, the strong outflow detected with Spitzer (Figure
\ref{spitzer_co}) is highly collimated, suggestive of a young source, but its
spatial extent suggests that the collapse of the driving source may
not have been as recent as that in IRAM 04191.  These characteristics
are similar to those of L483 which is thought to be in transition to a
Class I object \citep{taf00}.

\citet{lee05} note that the centers of the molecular outflow and
envelope surrounding IRAM 04191 are offset from the the continuum
source by approximately 560 AU. They suggest that the two cavities
seen to the south of that source could be due to a binary in which one
companion is too young to have any appreciable dust emission. There is
also evidence for binarity within B1c, particularly in the behavior of
its outflow ($\S$ \ref{res-outflow}), which shows evidence of
precession.  The cavity center is also offset from the continuum
source position.  There is as yet no detection of a second dust peak
nor a second outflow in B1c.

\subsection{Rotational Support}

The observed global rotation velocities within the B1 cloud are
insufficient to support the cloud against collapse by a factor of
$\sim 8$ \citep{bach90}.  The ages of embedded but optically visible
objects LkH$\alpha$ 327 and LkH$\alpha$ 328 are between 4-6 $\times
10^6$ yr \citep{ck79}.  Based on this, \citet{bach90} conclude that
another mechanism must be providing substantial support to the B1
cloud.  Within B1c, the two peaks of the rotating torus can be used to
derive the rotational energy relative to the $V_{LSR}$ at the core
center of B1c itself.

The inner edges of the torus are offset from the position of the
continuum peak.  For a sphere, the rotational energy is given by 
\begin{equation}
E_{\rm rot} = \frac{2}{5} \ \rho \ V \ v_r^2
\label{rotenergy}
\end{equation}

\noindent where $\rho$ is the mean density, $V$ is the volume, 
and $v_r$ is the radial velocity relative to the center.  Based on
estimates of the central density with \CEIGHTEENO\ ($\S$
\ref{disc-c18o}) and dust emission \citep{kir06}, we adopt a mean
density of $5 \times 10^5$ cm$^{-3}$ for B1c.

The radial velocity of the two peaks relative each other is $\sim
0.06$ \kms\ (as derived from Figure \ref{spectra}).  Thus, we estimate
the radial velocity relative to the core center to be a relatively
small shift of $\sim 0.03$
\kms.  The mean offset of the peaks from the core's velocity center is
$\sim 4$ \arcsec, or 1000 AU (for $d=250$ pc), encompassing a volume
of $1.4 \times 10^{49}$ cm$^3$.  Thus, based on equation
(\ref{rotenergy}), the rotational energy of the core is $\sim 1.2
\times 10^{39}$ ergs.

\citet{mw02} derived the energetics of the B1c core based on estimates
of the magnetic field dispersion and found rough equipartition between
the gravitational, kinetic and magnetic energies (all $\sim 10^{48}$
ergs).  The rotational energy is thus likely negligible to the support
of the core compared to the other means of support available through
nonthermal gas motions and magnetic field support.

The radius of the torus ($\sim 1000$ AU) detected in B1c leads
naturally to a discussion of a ``pseudo-disk'' \citep{gal93} around
the protostar within this core.  Such a disk is magnetically supported
and grows with time.  \citet{mw02} estimated the magnetic field
strength in the low density regime around the B1 cores to be on the
order of 30 $\mu$G.  While the field strength within the B1c core may
be higher, there is no Zeeman splitting measurement of sufficient
spatial resolution to isolate the field strength of the individual
cores within B1.  Assuming the presence of a pseudo-disk as described
by \citet{gal93}, we can use their equation (24) to estimate the age
of the disk around B1c.  For a field strength of 30 $\mu$G and a sound
speed of 0.35 \kms, a disk of radius 1000 AU has an age of $1.8 \times
10^5$ yr, consistent with the expected age of a Class 0 protostar.

\section{Summary}
\label{sum}

Utilizing interferometric and single dish millimeter data from the
BIMA array and FCRAO, we have probed the structure of the envelope of
the source Barnard 1c in Perseus.  We detect a powerful molecular
outflow in \co\ and \hcop\ which has carved a conical cavity into the 
envelope of B1c, destroying \NTWOHP.  The
prominence of the outflow driven by this source is also evident in
Spitzer IRAC data recently published by \citet{jor06}.  Comparison of
the CO molecular outflow and 4.5 \micron\ data reveals that positions
of changes in jet direction are mirrored by changes in velocity in the
CO emission.

We have detected significant \ntwohp\ emission from B1c.  Data from
FCRAO reveal \NTWOHP\ emission is of comparable scale to single dish
dust continuum maps.  High resolution data (created by uniform
weighting of the BIMA data combined with the FCRAO data) reveal
clearly a small cavity in which dense gas has been destroyed, most
likely by heating by the outflowing gas. We have detected a gradient
in $V_{LSR}$ of \NTWOHP\ consistent with rotation; moment maps reveal
evidence of a rotating torus offset from the continuum peak position.
B1c is similar to the more luminous source L483.  Both L483 and B1c
show C$^{18}$O toward the source peak and an anticorrelation of
C$^{18}$O with \NTWOHP, indicating that the interior is warm and that
CO has been desorbed from grains in that core interiors.  In a
subsequent paper (Matthews et al. 2006, in preparation), we will model
the centrally heated region within the B1c core using new data from
the SubMillimeter Array.

Based on the 3.3 mm dust continuum emission, we estimate that the mass
of the B1c inner envelope is in the range of $2.1 - 2.9 M_\odot$.
This is consistent with the mass of 2.4 $M_\odot$ derived from
the JCMT observations \citep{kir06} for a mean dust temperature
of 15 K.  However, the 3.3 mm measurement is only an estimate of the
inner envelope due to the spatial filtering of the BIMA array
data.

It is interesting to note that this core exhibits many features of the
standard picture for an evolving young stellar object.  In continuum
dust emission, the core has a dense centrally peaked interior.
However, the \ntwohp\ emission appears elongated along an orientation
roughly orthogonal to its outflow.  The two \NTWOHP\ peaks suggest
the presence of a rotating torus (with $r \sim 1000$ AU) with a central
cavity carved out by the outflow and heating by the central source.
We argue the heating has released \CO\ and its isotopes from grains
near the central source, destroying \NTWOHP.  \NTWOHP\ survives in the
rest of the inner envelope, as indicated by the torus (or pseudodisk)
and the \NTWOHP\ emission which brackets the outflow as traced in \co\
and \hcop.   As is the case in L483, the double peaks of \NTWOHP\
(torus) do not lie orthogonal to the outflow orientation.  In B1c,
the offset is $\sim 65$\degr. 

Existing polarimetry data \citep{mw02}, interpreted in the case of a
constant field orientation through the core would indicate a magnetic
field direction roughly parallel to the outflow orientation.  As yet,
there is no estimate of the field strength within B1c itself; however,
using the estimate from the surrounding B1 cloud \citep{mw02} and the
model of \citet{gal93}, the source age is estimated to have a lower
limit of $1.8 \times 10^5$ yr.  The dynamical age estimate of $3.7
\times 10^4$ yr is highly dependent on the inclination of the
outflow. 

In our forthcoming paper (Matthews et al., in preparation), we will
present higher transition observations of \CO\ and \CEIGHTEENO\ from
the B1c outflow, as well as measurements of the 1.3 mm continuum from
the SubMillimeter Array.  With existing data, and new data from IRS on
Spitzer, we will generate the spectral energy distribution and use it
to constrain models of $T(r)$ and $\rho(r)$ which was not possible
based on the weak continuum detections reported here.

\acknowledgements

The authors thank an anonymous referee for an insightful and thorough
report which improved the quality of the paper substantially.  We
would like to thank Mark Heyer for obtaining the FCRAO data for us
outside the normal proposal process.  We thank Paola Caselli for help
with the CLASS software and the hfs component fits and James Di
Francesco for many helpful discussions.  BCM's research was supported
by an NSERC PDF and Berkeley NSF grant 02-28963.  The research of MRH
was supported by a VIDI grant from the Nederlandse Organisatie voor
Wetenschappelijk Onderzoek.  The research of JKJ was supported by NASA
Origins Grant NAG 5-13050.  EAB acknowledges support from the NSF
grant 03-35207.  The BIMA array was operated with support from the
National Science Foundation under grants AST-02-28963 to UC Berkeley,
AST-02-28953 to U.\ Illinois, and AST-02-28974 to U.\ Maryland.  FCRAO
is supported by NSF Grant AST 02-28993.




\begin{thebibliography}{}

\bibitem[Aikawa et al.(2005)]{aik05}Aikawa, Y., Herbst, E., Roberts,
  H., \& Caselli, P. 2005, \apj, 620, 330

\bibitem[Arce \& Sargent(2006)]{arc06}Arce, H.G., \& Sargent,
  A.I. 2006, \apj, 646, 1070

\bibitem[Andr\'{e}, Motte \& Bacmann(1999)]{and99}Andr\'{e}, P., Motte,
  F., \& Bacmann, A. 1999, \apj, 513, L57

\bibitem[Bachiller \& Cernicharo(1986)]{bc86a}Bachiller, R., \&
Cernicharo, J. 1986, \aap, 166, 283

\bibitem[Bachiller \etal(1990)]{bach90}Bachiller, R., Menten, K.M., \&
del Rio-Alvarez, S. 1990, \aap, 236, 461

\bibitem[Bachiller et al.(2001)]{bac01} Bachiller, R., 
P{\'e}rez Guti{\'e}rrez, M., Kumar, M.~S.~N., \& Tafalla, M.\ 2001,
\aap, 372, 899 

\bibitem[Beckwith, Henning \& Nakagawa(2000)]{bec00}Beckwith, S.V.W.,
  Henning, Th., \& Nakagawa, Y. 2000, Protostars \& Planets IV, 533

\bibitem[Belikov \etal(2002)]{beli02}Belikov, A.N., Kharchenko, N.V.,
  Piskunov, A.E., Shilbach, E., \& Scholz, R.-D. 2002, \aap, 387, 117

\bibitem[Belloche \& Andr\'{e}(2004)]{bel04}Belloche, A., \& Andr\'{e},
  P. 2004, \aap, 419, L35

\bibitem[Belloche, Andr\'{e}, Despois \&
  Blinder(2002)]{bel02}Belloche, A., Andr\'{e}, Despois, D., \&
  Blinder, S. 2002, \aap, 927, 947

\bibitem[Bergin \& Langer(1997)]{ber97}Bergin, E.A., \& Langer,
  W.D. 1997, \apj, 486, 316

\bibitem[Bergin et al.(1998)]{bnm98} Bergin, E.~A., Neufeld, 
D.~A., \& Melnick, G.~J.\ 1998, \apj, 499, 777 

\bibitem[Bergin \etal(2002)]{ber02}Bergin, E.A., Alves, J., Huard, T.,
  \& Lada, C.J. 2002, \apj, 570, L101

\bibitem[Bisschop \etal(2006)]{bis06}Bisschop, S.E., Fraser, H.J.,
  \"{O}berg, K.I., van Dischoek, E.F., \& Schlemmer, S. 2006, \aap,
  449, 1297

\bibitem[Bontemps \etal(1996)]{bon96}Bontemps, S., Andr\'{e}, P.,
  Tereby, S., Cabrit, S. 1996, \aap, 311, 858

\bibitem[Borgman \& Blaauw(1964)]{bb64}Borgman, J., \& Blaauw,
A. 1964, Bull.\ Astron.\ Inst.\ Netherlands, 17, 358

\bibitem[Buisson \etal(2002)]{bui02}Buisson, G., Desbats, L., Duvert,
  G., Forveille, T., Gras, R., Guilloteau, S., Lucas, R., \& Valiron,
  P. 2002, Continuum and Line Analysis Single-Dish System Manual
  (IRAM: Grenoble), http://iram.fr/GS/class/class.html

\bibitem[{\u C}ernis(1990)]{cer90}{\u C}ernis, K. 1990, \apss, 166,
  315

\bibitem[{\u C}ernis \& Strai{\u z}ys(2003)]{cer03}{\u C}ernis, K., \&
  Strai{\u z}ys, V. 2003, Baltic Astronomy, 2, 214

\bibitem[Caselli \etal(2002)]{cas02}Caselli, P., Benson, P., Myers,
  P., \& Tafalla, M. 2002, \apj, 572, 238

\bibitem[Cernicharo \etal(1985)]{cer85}Cernicharo, J., Bachiller, R.,
\& Duvert, G. 1985, \aap, 149, 273

\bibitem[Cohen \& Kuhi(1979)]{ck79}Cohen, M., \& Kuhi, L.V. 1979, \apjs,
41, 743

\bibitem[de Zeeuw, Hoogerwerf \& de Bruijne(1999)]{dez99}de Zeeuw, P.T., Hoogerwerf, R., \&
  de Bruijne, J.H.J. 1999, \aj, 117, 354

\bibitem[Di Francesco \etal(2006)]{dif06}Di Francesco, J., Evans,
  N.J.,II, Caselli, P., Myers, P.C., Shirley, Y., Aikawa, Y., \&
  Tafalla, M., 2006, in ``Protostars \& Planets V,'' eds. B.\
  Reipurth, D.\ Jewitt, \& K.\ Kiel, (University of Arizona Press:
  Tucson), in press

\bibitem[Dore \etal(2003)]{dor03}Dore, L., Caselli, P., Beninati, S.,
  Bourke, T., Myers, P.C., \& Cazzoli, G. 2003, \aap, 413, 1177

\bibitem[Enoch \etal(2006)]{eno06}Enoch, M.L., \etal 2006, \apj, 638,
  293

\bibitem[Frerking, Langer \& Wilson(1982)]{fre82}Frerking, M.A.,
  Langer, W.D., \& Wilson, R.W. 1982, \apj, 262, 590

\bibitem[Galli \& Shu(1993)]{gal93}Galli, D., \& Shu, F.H. 1993, \apj,
  417, 243

\bibitem[Gooch(1996)]{goo96}Gooch, R.E. 1996, ADASS, ASP Conf. Series
101, ed. G.H. Jacoby \& J. Barnes, APS, San Francisco, 80-83

\bibitem[Herbig \& Jones(1983)]{her83}Herbig, G.H., \& Jones,
  B.F. 1983, \aj, 88, 1040

\bibitem[Hirano \etal(1997)]{hir97}Hirano, N., Kameya, O., Mikami, H.,
Saito, S., Umemoto, T., \& Yamamoto, S.  1997, \apj, 478, 631

\bibitem[Hodapp \etal(2005)]{hod05}Hodapp, K.W., Bally, J.,
  Eis\"{o}ffel, J., \& Davis, C.J. 2005, \aj, 129, 1580

\bibitem[Hogerheijde(2005)]{hog05}Hogerheijde, M.R. 2005, IAU Symposium, 231, 90

\bibitem[J{\o}rgensen(2004)]{jor04}J{\o}rgensen, J.K. 2004, \aap, 424, 589

\bibitem[J{\o}rgensen \etal(2006)]{jor06}J{\o}rgensen, J.K., \etal 2006,
    \apj, 645, 1246

\bibitem[J{\o}rgensen \etal(2004a)]{jor04a}J{\o}rgensen, J.K.,
  Hogerheijde, M.R., Blake, G.A., van Dishoeck, E.F., Mundy, L.G., \&
  Sch\"{o}ier, F.L. 2004a, \aap, 415, 1021

\bibitem[J{\o}rgensen, Sch\"{o}ier \& van Dishoeck (2004b)]{jor04b}J{\o}rgensen, J.K.,
  Sch\"{o}ier, F.L., \& van Dishoeck, E.F.  2004b, \aap, 416, 603

\bibitem[J{\o}rgensen, Sch\"{o}ier \& van Dishoeck (2002)]{jor02}J{\o}rgensen, J.K.,
  Sch\"{o}ier, F.L., \& van Dishoeck, E.F.  2002, \aap, 389, 908

\bibitem[Kirk, Johnstone \& Di Francesco(2006)]{kir06}Kirk, H.,
  Johnstone, D., \& Di Francesco, J. 2006, \apj, in press

\bibitem[Lee, Ho \& White(2005)]{lee05}Lee, C.-F., Ho, P.T.P., \&
  White, S.M. 2005, \apj, 619, 948

\bibitem[Lee et al.(2004)]{lbe04}Lee, J.-E., Bergin, E.~A., 
\& Evans, N.~J.\ 2004, \apj, 617, 360 

\bibitem[Lynds(1969)]{l69}Lynds, B.T., 1969, \pasp, 81, 496

\bibitem[Matthews \& Wilson(2002)]{mw02}Matthews, B.C., \& Wilson, 
C.D. 2002, ApJ, 577, 864

\bibitem[Matthews(2005)]{mat05}Matthews, B.C. 2005, in the proceedings
  of ``Astronomical Polarimetry: Current Status and Future
Directions'', eds.\ Adamson A., Davis C.J., Aspin
  C., ASP Conference Series, 343, 99

\bibitem[Noriega-Crespo et al.(2004)]{nor04}Noriega-Crespo, A., et
  al. 2004, \apjs, 154, 352

\bibitem[\"{O}berg et al.(2005)]{obe05}\"{O}berg, K.I., van
  Brokhuizen, F., Fraser, H.J., Bisschop, S.E., van Dischoek, E.F., \&
  Schlemmer, S. 2005, \apj, 621, L33

\bibitem[Ossenkopf \& Henning(1994)]{oss94}Ossenkopf, V., \& Henning,
  Th. 1994, \aap, 291, 943

\bibitem[Pirogov \etal(2003)]{pir03}Pirogov, L., Zinchenko, I.,
  Caselli, P., Johansson, L.E.B., \& Myers, P.C. 2003, \aap, 405, 639

\bibitem[Ridge \etal(2006)]{rid05}Ridge, N. et al. 2006, AJ, 131, 2921

\bibitem[Sargent(1979)]{sar79}Sargent, A.I., 1979, \apj, 233, 163

\bibitem[Sault, Teuben \& Wright(1995)]{sau95}Sault, R.J., Teuben,
  P.J., \& Wright, M.C.H. 1995, ASPC, 77, 433

\bibitem[Sch\"{o}ier \& van der Tak(2005)]{sho05}Sch\"{o}ier, F.L., van
  der Tak, F.F.S., van Dishoeck, E.F., Black, J.H. 2005, \aap, 432, 369

\bibitem[Stanimirovic et al.(1999)]{sta99}Stanimirovic, S.,
  Staveley-Smith, L., Dickey, J.M., Sault, R.M., \& Snowden, L. 1999,
  MNRAS, 302, 417 

\bibitem[Tafalla \etal(2000)]{taf00}Tafalla, M., Myers, P.C.,
  Mardones, D., \& Bachiller, R. 2000, \aap, 359, 967

\bibitem[Welch \etal(1996)]{wel96}Welch, J.W., \etal 1996, \pasp, 108, 93

\bibitem[Whittet \etal(2001)]{whi01}Whittet, D., Gerakines, P., Hough,
  J., \& Shenoy, S. 2001, \apj, 547, 872

\bibitem[Whitworth \& Ward-Thompson(2001)]{ww01}Whitworth, A.P., \&
  Ward-Thompson, D. 2001, \apj, 547, 317

\bibitem[Williams, de Geus \& Blitz(1994)]{wil94}Williams, J.P., de
  Geus, E.J., \& Blitz, L. 1994, \apj, 428, 693

\bibitem[Young \etal(2004)]{you04}Young, C.H. et al. 2004, \apjs, 154, 396

\end{thebibliography}
\end{document}